\documentclass[prd,preprintnumbers,pacs,12pt,nofootinbib]{revtex4}
\usepackage{amsmath}
\usepackage{amssymb}
\usepackage{overpic,color}
\usepackage{graphicx} 
\usepackage{epstopdf} 

\usepackage{bm}
\usepackage{amsmath}
\usepackage{amssymb}

\usepackage[latin1]{inputenc}
\usepackage{slashed}
\usepackage{amscd}
\usepackage{latexsym}
\begin{document}

\newcommand{\lag}{\mathcal{L}}
\newcommand{\de}{\partial}
\newcommand{\To}{\Rightarrow}
\newcommand{\bra}[1]{\langle #1|}
\newcommand{\ket}[1]{|#1\rangle}
\newcommand{\vev}[1]{\langle #1 \rangle}
\newcommand{\tr}[1]{\textrm{Tr}\big[#1\big]}
\newcommand{\alg}[1]{\mathfrak{#1}}
\newcommand{\nda}[1]{\emph{(#1 - n.d.a.)}}
\newcommand{\gf}[1]{\mathbf{#1}}


\title{From hidden symmetry to extra dimensions: a five dimensional formulation of the Degenerate BESS model}

 \author{Francesco Coradeschi, Stefania De Curtis and Daniele Dominici}
 \affiliation{Department of Physics, University of Florence, and
 INFN,  Via Sansone 1, 50019 Sesto F., (FI), Italy}

\date{\today}

 \begin{abstract}
\noindent
We consider the continuum limit of a moose model corresponding to a generalization to $N$ sites of the Degenerate BESS model. The five dimensional formulation emerging in this limit is a realization of a RS1 type model with $SU(2)_L\otimes SU(2)_R$ in the bulk, broken by boundary conditions and a vacuum expectation value on the infrared  brane. A low energy effective Lagrangian is derived by means of the holographic technique and corresponding bounds on the model parameters are obtained.
 \end{abstract}
 \pacs{12.60.Cn, 11.25.Mj, 12.39.Fe}

 \maketitle


\section{Introduction}

The exact nature of the mechanism that leads to the breakdown of the electroweak 
(EW) symmetry  is one of the relevant open questions in particle physics. 
 While waiting for the first experimental data from the Large Hadron Collider, it is worthwhile to explore the potential electroweak  breaking scenarios from a theoretical point of view.

In the Standard Model (SM), the mechanism of the EW symmetry breaking implies the presence a fundamental scalar particle, the Higgs boson, with a light mass as suggested by EW fits. However, this mechanism is affected by a serious fine-tuning problem, the hierarchy problem, because the mass of the Higgs boson is not protected against radiative corrections and would naturally be expected to be as large as the physical UV cut-off of the SM, which could be as high as $M_P \simeq 10^{19}$ GeV.

Possible solutions to the hierarchy problem  are the technicolor (TC) theories \cite{Weinberg:1975gm,Susskind:1978ms,Weinberg:1979bn} (that postulate the presence of new strong interactions around the TeV scale) and extra-dimensional theories \cite{ArkaniHamed:1998rs,Randall:1999ee,Randall:1999vf}. These seemingly unrelated classes of theories have in fact a profound connection through the AdS/CFT correspondence \cite{Maldacena:1997re}.
 According to this conjecture, five dimensional (5D) models on  AdS space are ``holographic duals'' to 4D theories with spontaneously broken conformal invariance. 
The duality means that  when the 5D theory is in a perturbative regime the holographic dual is strongly interacting and vice versa.
 This fact provides an unique tool to make quantitative calculations in 4D strongly interacting theories, and creates a very interesting connection between extra-dimensional and TC-like theories.

Working in the framework suggested by the
AdS/CFT correspondence, one considers for  the fifth dimension  a segment ending
with two branes (one ultraviolet, UV, and the other one infrared, IR). Several choices of gauge groups in the bulk have been proposed: there are models, with or without the Higgs,  which assume a $SU(2)_L\times SU(2)_R\times
U(1)_{B-L}$ gauge group, 
\cite{Csaki:2003dt,Agashe:2003zs,Csaki:2003zu,Cacciapaglia:2004zv,Cacciapaglia:2004rb,Cacciapaglia:2004jz,Davoudiasl:2004pw,Carena:2006bn,Cacciapaglia:2006mz,Contino:2006nn}, or $SU(2)\times U(1)$ \cite{Carena:2002me,Csaki:2002gy,Carena:2002dz,Cui:2009dv}
or a simpler  $SU(2)$   in the 5D bulk
\cite{Foadi:2005hz,Casalbuoni:2007xn}. Compactification from five to four dimensions is often performed by the standard Kaluza Klein mode expansion, however alternative methods have been suggested.
Effective low energy chiral Lagrangians in
four dimensions, can be obtained by
 holographic versions of 5D theories in warped
background \cite{Nomura:2003du,Barbieri:2003pr,Barbieri:2004qk}
or via 
the deconstruction technique
\cite{ArkaniHamed:2001ca,ArkaniHamed:2001nc,Hill:2000mu,Cheng:2001vd,Abe:2002rj,Falkowski:2002cm,Randall:2002qr,Son:2003et,deBlas:2006fz}.
The deconstruction mechanism provides  a correspondence at
low energies between theories with  replicated 4D
gauge symmetries $G$   and theories with a 5D gauge
symmetry $G$ on a lattice. 

We will refer to the deconstructed models also  as  moose models. Several examples mainly based on the gauge group $SU(2)$
 have recently received attention \cite{Foadi:2003xa,Foadi:2004ps,Hirn:2004ze,Casalbuoni:2004id,Chivukula:2004pk,Georgi:2004iy,Casalbuoni:2005rs,Bechi:2006sj,Coleppa:2006fu}. It is interesting to note
that even few sites of the moose give a good approximation of the 5D
theory \cite{Becciolini:2009fu}.  The BESS model \cite{Casalbuoni:1985kq,Casalbuoni:1986vq}, based on the hidden symmetry approach \cite{Bando:1984ej,Bando:1987br,Bando:1987ym}, is a prototype of models of this kind: with a particular choice of its
parameters it can generate the recently investigated three site model
\cite{Chivukula:2006cg,Matsuzaki:2006wn,Matsuzaki:2006zz,Dawson:2007yk,Abe:2008hb}. 

Generic TC models  usually have difficulties in satisfying the constraints coming from EW precision measurements on $S,T,U$ observables \cite{Peskin:1990zt,Peskin:1991sw}.
In this paper, we have reconsidered  the Degenerate BESS (D-BESS) model \cite{Casalbuoni:1995yb,Casalbuoni:1995qt}, a low-energy effective  theory which possesses a $\left(SU(2) \otimes SU(2)\right)^2$ custodial symmetry  that leads to a suppressed contribution from the new physics to the EW precision observables, making  possible to have new vector bosons at a relatively low energy scale (around a TeV). This new vector states are interpreted as composites of a strongly interacting sector.
Starting from   the generalization to $N$ sites of the D-BESS model (GD-BESS) \cite{Casalbuoni:2004id,Casalbuoni:2007dk}, based on the deconstruction  or ``moose'' technique \cite{ArkaniHamed:2001ca}, we consider  the continuum limit to  a 5D theory.

The D-BESS model and its generalization suffer the drawback of the unitarity constraint, which is as low as that of the Higgsless SM \cite{Appelquist:1980vg,Longhitano:1980iz}, that is around $1.7$ TeV. However, (at least for a particular choice of the extra-dimensional background) in the 5D model it is meaningful to reintroduce an Higgs field, delaying unitarity violation to a scale $\gtrsim 10$ TeV. In the ``holographic'' interpretation of AdS$_5$ models \cite{ArkaniHamed:2000ds,Rattazzi:2000hs}, inspired by the AdS/CFT correspondence, this 
Higgs can be thought as a composite state and thus does not suffer from the hierarchy problem. The 5-dimensional D-BESS (5D-DBESS) on AdS$_5$ then provides a coherent description of the low energy phenomenology of a new strongly interacting sector up to energies significantly beyond the $\sim 2$ TeV limit of the Higgsless SM, still showing a good compatibility with EW precision observables. 
 While studying this 5D extension, furthermore, we have clarified a not-so-obvious fact: there is at least a particular limit, where the 5D-DBESS can be related to a realization of the RS1 model \cite{Randall:1999ee}, specifically the one proposed in ref. \cite{Carena:2002dz}.

In Section \ref{review} we review the generalization of D-BESS to $N$ sites. In Section \ref{tofiveD} we extend the model to five dimensions clarifying how the boundary conditions at
 the ends of the 5D segment emerge from the deconstructed version of the model. The 5D model is described by  a $SU(2)_L\times SU(2)_R$ bulk Lagrangian with boundary kinetic terms, broken both spontaneously and by boundary conditions. In Section \ref{expansion} we develop the expansion in mass eigenstates consisting in  two charged gauge sectors, left (including $W$) and right, and one neutral (including the photon and the $Z$). We also  get  the expanded Lagrangian for the modes. In Section \ref{lowen5D}  we derive the low energy Lagrangian by means of the holographic technique and the EW precision parameters 
$\epsilon_{1,2,3}$. In Section \ref{phen}  we show  the spectrum of KK 
excitations and derive the bounds from EW precision measurements on the model parameters for two choices of the 5D metric, the flat case and the RS one. 
Conclusions are given in Section \ref{conclusions}. In Appendix \ref{appA},  by following 
\cite{Falkowski:2006vi,Falkowski:2007iv},
 we develop the technique for Kaluza-Klein expansion.


\section{Review of the GD-BESS model}
\label{review}

\begin{figure}
\includegraphics[width=\textwidth]{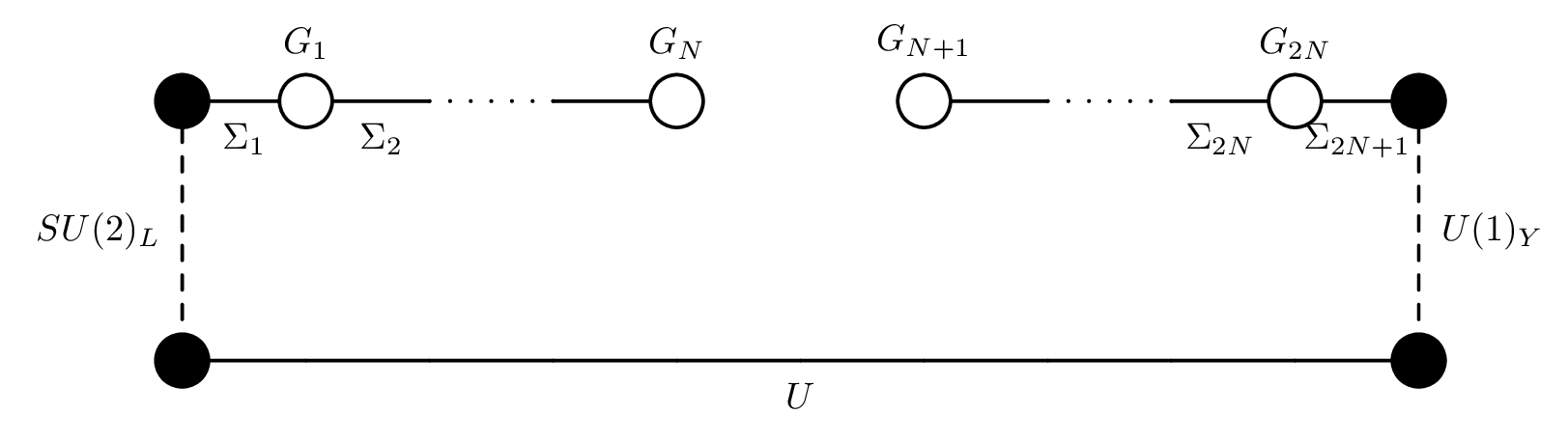}
\caption{\small{{ \it Graphic representation of the moose model described by the Lagrangian given in Eq.~\eqref{eq:1}. The dashed lines represent the identification of the corresponding moose sites.}}}
\label{fig:A}
\end{figure}

The GD-BESS model is a moose model with a $\left[SU(2)\right]^{2N} \otimes SU(2)_L \otimes U(1)_Y$ gauge symmetry, described by the Lagrangian \cite{Casalbuoni:2007dk}:
\begin{equation}
\begin{split}
\label{eq:1}
\mathcal{L} & =\sum_{i=1, i \neq N+1}^{2N+1} f_i^2 \textrm{Tr}[D_\mu\Sigma_i^\dag D^\mu \Sigma_i]  + f_0^2 \textrm{Tr} [D_\mu U^\dagger D^\mu U]\\
& - \frac{1}{2 \tilde{g}^2}\textrm{Tr}[(\mathbf{F}_{\mu \nu}^{\tilde{W}})^2]- \frac{1}{2 \tilde{g}^{'2}}\textrm{Tr}[(\mathbf{F}_{\mu \nu}^{\tilde{B}})^2]- \frac{1}{2 g_i^2}\sum_{i=1}^{2N}\textrm{Tr}[(\gf{F}_{\mu \nu}^i)^2],
\end{split}
\end{equation}
where 
\begin{equation}
\gf{F}_{\mu \nu}^i = \de_\mu \gf{A}_\nu^i -  \de_\nu \gf{A}_\mu^i + i [\gf{A}_\mu, \gf{A}_\nu], \quad i = 0, \ldots 2N+1
\end{equation}
with
$\gf{A}^i \equiv A^{a \, i} \frac{\tau^a}{2}, i = 1, \ldots 2N$ are the $\left[SU(2)\right]^{2N}$ gauge fields, $\gf{A}^0 \equiv\gf{\tilde{W}} \equiv \tilde{W}^{a} \frac{\tau^a}{2}$ and $\gf{A}^{2N+1} \equiv \gf{\tilde{B}} \equiv \tilde{B} \frac{\tau^3}{2}$ are the $SU(2)_L \otimes U(1)_Y$ gauge fields,
and the chiral fields $\Sigma_i$ and $U$ have covariant derivatives defined by
\begin{align}
D_\mu \Sigma^i = \de_\mu \Sigma^i + i & \gf{A}_\mu^{i-1} \Sigma^i - i \Sigma^i \gf{A}_\mu^i, \quad i = 1, \ldots 2N+1, \ i \neq N+1\\
\label{covderU}
& D_\mu U = \de_\mu U + i \gf{\tilde W}_\mu \, U - i U \, \gf{\tilde B}_\mu
\end{align}
 The model has two sets of parameters, the gauge coupling constants $\tilde{g}, \ \tilde{g}', g_i, \ i=1, \ldots 2N$, and the ``link coupling constants'' $f_i, \ i = 1, \ldots 2N+1, \ i \neq N+1$. For simplicity, we will impose a reflection invariance with respect to the ends of the moose, getting the following relations among the model parameters
\begin{equation}
\label{reflection}
f_i = f_{2N+2-i}, \quad g_i = g_{2N+1-i}.
\end{equation}
The model content in terms of fields and symmetries can be summarized by the moose in Fig.~ \ref{fig:A}.

The most peculiar feature is the absence in the Lagrangian given in Eq.~\eqref{eq:1} of any field connecting the $N^{th}$ and the ${N+1}^{th}$ sites of the moose. This situation was referred to in ref. \cite{Casalbuoni:2004id} as ``cutting a link''. As it was shown there, this choice guarantees the vanishing of the leading order corrections from new physics to the EW precision parameters; for instance, in ref. \cite{Casalbuoni:2007dk}, the contributions to the $\epsilon$ parameters \cite{Altarelli:1990zd,Altarelli:1991fk,Altarelli:1993sz} as well as the ``universal'' form factors \cite{Barbieri:2004qk} were  calculated, showing that all these contributions are of order $m_Z^2/\bar{M}^2$, where $\bar{M}$ is the mass scale of the lightest new resonance in the theory.

\section{Generalization to 5 dimensions}
\label{tofiveD}

We now want to describe the continuum limit $N \to \infty$ of the Lagrangian given in Eq.~\eqref{eq:1}. As it is well known, a $\left[SU(2)\right]^K$ linear moose model can be interpreted as the discretized version of a SU(2) 5D gauge theory. The GD-BESS model, however, has a number of new features with respect to a basic linear moose. In particular, we have the ``cut link'' and the presence of an apparently nonlocal field $U$ which connects the gauge fields of the $SU(2)_L \otimes U(1)_Y$ local symmetry.

To be able to properly describe the 5D generalization, we need a representation for the 5D metric. Since the deconstructed model possesses ordinary 4D Lorentz invariance, the extra-dimensional metric must be compatible with this symmetry. Such a metric can in general be written in the form:
\begin{equation}
ds^2 = b(y) \eta_{\mu \nu} \, dx^\mu dx^\nu + dy^2,
\end{equation}
where $\eta$ is the standard Lorentz metric with the $(-,+,+,+)$ signature choice, $y$ the variable corresponding to the extra dimension and $b(y)$ is a generic positive definite function, usually known as the ``warp factor''. We normalize $b(y)$ by requiring that $b(0) = 1$. For definiteness, we will consider a finite extra dimension, with $y \in (0, \pi R)$. With this choice, it is possible to write down an identification between the GD-BESS and the continuum limit parameters:
\begin{equation}
\label{identification}
\frac{g_i^2}{N} \to \frac{g_5^2}{\pi R}, \quad f_i^2 \to b(y) \frac{N}{\pi R g_5^2},
\end{equation}
where $g_5$ is a 5D gauge coupling, with mass dimension $-1/2$. As can be seen, a general choice for the $g_i$ implies that $g_5$ is ``running'' , with an explicit dependence on the extra variable. In the following, we will not consider this possibility, but rather restrict for simplicity to a constant coupling (as it is standard in the literature), so, from the 4D side, we will have $g_i \equiv g_c$.

\begin{figure}
 \begin{center}
 \includegraphics[width=0.9\textwidth]{5Dmoose1.eps}
  \parbox{\textwidth}{\huge{$\mspace{0mu}\Downarrow$}}
  \includegraphics[width=\textwidth]{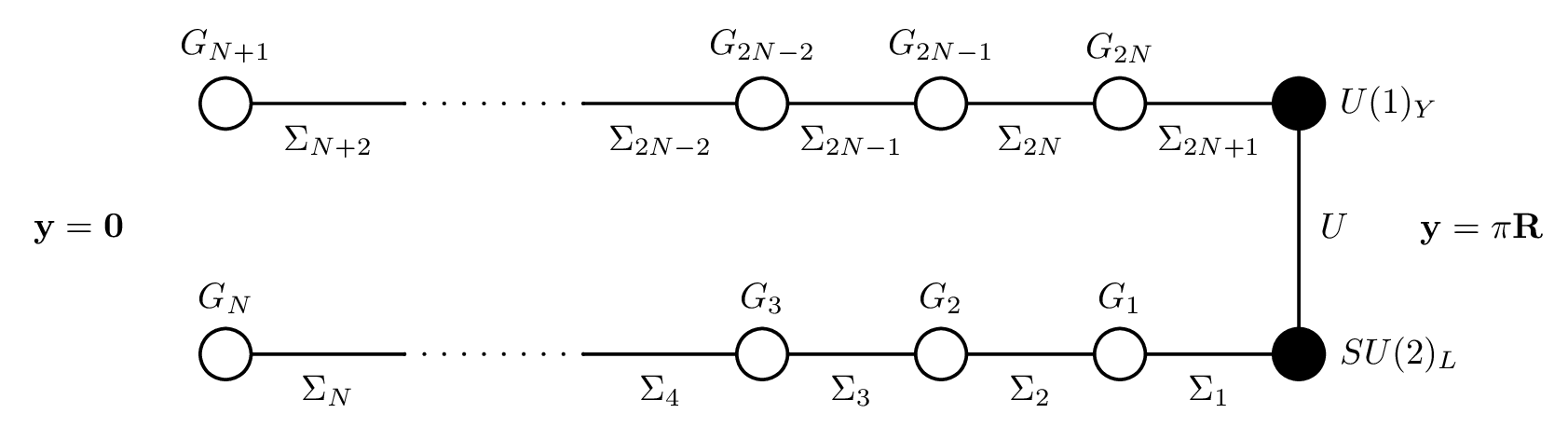}
  \caption{\small{\textit{Interpretation of the cut link in the continuum limit of the GD-BESS model. The first half of the moose is ``flipped'' and superimposed to the second half. In this way, the $N^{th}$ and the ${N+1}^{th}$ sites are identified with the $y=0$ brane, while the $1^{st}$ and the ${2N+1}^{th}$ with the $y = \pi R$ one.}}}
\label{moose5Df}
 \end{center}
 \end{figure}

The trickiest part of the generalization, however, is to interpret the cutting of the link. To understand this properly, we can start by noticing that the cut link prevents any direct contact between the two sides of the moose; the fields on the left only couple to those on the right through the field $U$. In this sense, the moose is split by the cut in two separate pieces, linked by $U$.  Due to the reflection symmetry (see Eq.~\eqref{reflection}), the two pieces are identical to each other, at a site-by-site level, from every point of view: field content, coupling constants $g_i$, link couplings $f_i$. The right way to look at this set up is to \emph{consider the sites connected by the reflection symmetry as describing the same point along the extra dimension}: for example, we can look at the fields $\gf{A}^i$ and $\gf{A}^{2N-i}$ not as values of the same 5D $SU(2)$ gauge field at two different points along the extra dimension, but rather \emph{components} of a single $SU(2) \otimes SU(2)$ gauge field at the \emph{same} extra-dimensional location. We can do this, because by Eq.~\eqref{identification}, the warp factor - and thus the 5D metric - at a given site, only depends on the value of the link coupling constant $f_i$, which is equal at points identified by the reflection symmetry, that  describe the same point $y_i$ on the $5^{th}$ dimension. The situation is depicted graphically in Fig.~ \ref{moose5Df}: it is equivalent to flip one of the pieces of the moose and superposing it to the other one. In this way, we do not obtain a 5D $SU(2)$ gauge theory, but a $SU(2)_L \otimes SU(2)_R$ one, with the left part of the moose describing the $SU(2)_L$ gauge theory and the right part $SU(2)_R$ and the coupling constants of the two sectors of the gauge group identified by a discrete symmetry. The field $U$ no longer appears as nonlocal, but rather as confined at one end of the extra-dimensional segment.

The last point to consider is the presence of different gauge fields - the ones corresponding to $SU(2)_L \otimes U(1)_Y$ - at the two ends of the moose, which are identified with one of the endpoints of the 5D interval (which for definiteness we will take to be $y = \pi R$). This can be accounted for by considering \emph{localized kinetic terms} at $y = \pi R$ for the 5D gauge fields; the fields $\tilde{W}$ and $\tilde{B}$ can then be simply identified with the values of the $SU(2)_L$ and of the third component of the $SU(2)_R$ 5D gauge fields respectively. Notice that the ``flipped'' GD-BESS moose has $N+1$ sites: $N$ for the $SU(2)_L \otimes SU(2)_R$ gauge fields and the last one for the fields corresponding to $SU(2)_L \otimes U(1)_Y$. By convention, we will map this last site to the $y = \pi R$ end of the extra dimension; the other endpoint, $y = 0$, will correspond to the gauge fields living next to the cut link, $\gf{A}^N$ and $\gf{A}^{N+1}$.

Putting all  together, the 5D limit of GD-BESS is described by the action
\begin{equation}
\label{5Daction}
\begin{split}
& S = \int d^4x \int_0^{\pi R} \sqrt{-g} \ dy \left[  - \frac{1}{4g_5^2}L_{M N}^a L^{a \; M N} - \frac{1}{4g_5^{2}}R_{M N}^a R^{a \; M N}\right. \\
+ \, \delta(y-\pi R) & \left( \left. - \frac{1}{4\tilde{g}^2} L_{\mu \nu}^a L^{a \; \mu \nu} - \frac{1}{4 \tilde{g}^{'2}}R_{\mu \nu}^3 R^{3 \; \mu \nu} - \, \frac{\tilde{v}^2}{4}(D_\mu U)^\dag D^\mu U + \text{\textit{fermion terms}} \right) \right],
\end{split}
\end{equation}
where:
\begin{itemize}
\item with the usual convention, the greek indices run from 0 to 3, while capital latin ones take the values $(0,1,2,3,5)$, with ``$5$'' labelling the extra direction
\item $g$ is the determinant of the metric tensor $g_{M N}$, defined by
\begin{equation}
ds^2 = g_{M N} dx^M dx^N \equiv b(y) \; \eta_{\mu \nu} d x^\mu d x^\nu + dy^2,
\end{equation}
\item $L_{M N}^a$ and $R_{M N}^a$ are the $SU(2)_L \otimes SU(2)_R$ gauge field strengths:
\begin{equation}
L(R)_{M N}^a = \de_M W_{L(R) \; N}^a - \de_N W_{L(R) \; M}^a + i \epsilon^{a b c} W_{L(R) \; M}^b W_{L(R) \; N}^c;
\end{equation}
the fields $W_{L(R)}^a$ represent the continuum limit of the $A^{a \, i}$
\item $\tilde{g}$, $\tilde{g}'$, $g_5$ are three, in general different, gauge couplings. $\tilde{g}$ and $\tilde{g}'$ are the direct analogous of their deconstructed counterparts. $g_5$ is the bulk coupling, it has mass dimension $-\frac{1}{2}$, and it is the 5D limit of the $g_i$, as can be seen by Eq.~\eqref{identification}
\item the brane scalar $U$ is a $SU(2)$-valued field, with its covariant derivative defined by:
\begin{equation}
D_\mu U = \de_\mu U + i W_{L \; \mu}^a \frac{\tau^a}{2} U - i W_{R \; \mu}^3 U \frac{\tau^3}{2},
\end{equation}
in exact analogy with Eq.~\eqref{covderU}. Note that the field $U$ is analogous to the one that describes the standard Higgs sector in the limit of an infinite Higgs mass \cite{Appelquist:1980vg}. It can be conveniently parametrized in terms of three real pseudo-scalar fields $\pi^a$,
\begin{equation}
U = \exp(\dfrac{i  \pi^a \tau^a}{2 \tilde{v}});
\end{equation}
\item the fermionic terms, which we take to be confined on the brane for simplicity, have the usual SM form.
\end{itemize}

It is important to notice that the action \eqref{5Daction} does not define the physics of the model uniquely: we still have the freedom of choosing \emph{boundary conditions} (BCs) for the fields. 
This BC ambiguity is absent in deconstructed models: the BCs get implicitly specified by the way in which the discretization of the $5^{th}$ dimension is realized. This means that the GD-BESS model described by Eq.~\eqref{eq:1} already has a specific set of ``built-in'' BCs. These can be understood by looking at the residual gauge symmetry at the ends of the moose. It is apparent that, after the ``flipping'' depicted in Fig.~ \ref{moose5Df}, at the $N^{th}$ and $(N+1)^{th}$ sites, corresponding to $y = 0$ in the continuum limit, we have the full $SU(2)_L \otimes SU(2)_R$ gauge invariance. By contrast, at the $0^{th}$ and $(2N+1)^{th}$ sites, corresponding to $y = \pi R$, the gauge symmetry is broken down to $SU(2)_L \otimes U(1)_Y$. To do this, we have to impose Dirichlet BCs on two of the $SU(2)_R$ gauge fields at $y = \pi R$, while all the other fields, and all the fields at $y = 0$ are unconstrained. The complete gauge symmetry breaking pattern is thus as follows: we have a $SU(2)_L \otimes SU(2)_R$ gauge invariance in the bulk, unbroken on the $y = 0$ brane and broken by a combination of Dirichlet BCs and scalar VEV (of the $U$ field) to $U(1)_{e.m.}$ on the $y = \pi R$ brane.

In the following of this work, we will study the model defined by the action \eqref{5Daction}. First of all, we will perform a general analysis of the full 5D theory by the standard technique of the Kaluza-Klein (KK) expansion. Then we will look at the low-energy limit and derive expression for the $\epsilon$ parameters; the results will confirm that this is indeed the 5D limit of GD-BESS. Finally, we will make some remarks on the phenomenology of the model in correspondence with two interesting choices for the geometry of the $5^{th}$ dimension, that of a flat dimension ($b(y) \equiv 1$) and that of a slice of AdS$_5$ ($b(y) = e^{-2k y}$).

\section{Expansion in mass eigenstates}
\label{expansion}

Since we wish to keep the metric generic for the moment, a convenient strategy is to expand the gauge fields $W_{L(R) \, M}^a$ (and the goldstones $\pi^a$) directly in terms of mass eigenstates \cite{Falkowski:2006vi,Falkowski:2007iv}. So we define:
\begin{equation}
\label{KKexp}
\begin{split}
W_{L \; \mu}^a (x,y) = \sum_{j = 0}^\infty f_{L \, j}^a (y) & \, V_\mu^{(j)}(x), \qquad W_{L \; 5}^a (x,y) = \sum_{j = 0}^\infty g_{L \, j}^a (y) \, G^{(j)}(x),\\
W_{R \; \mu}^a (x,y) = \sum_{j = 0}^\infty f_{R \, j}^a (y) & \, V_\mu^{(j)}(x), \qquad W_{R \; 5}^a (x,y) = \sum_{j = 0}^\infty g_{R \, j}^a (y) \, G^{(j)}(x),\\
& \pi^a (x) = \sum_{j = 0}^\infty c_j^a \, G^{(j)}(x).
\end{split}
\end{equation}


The expansion \eqref{KKexp} is written in full generality. \emph{A priori}, this means that fields with different $SU(2)$ index could be mixed. The index ``$(j)$'' labels all the mass eigenstates. This procedure is in fact more general than it is needed; we will see that, with our choice  of BCs for the model, we will get  three decoupled towers of eigenstates, so that many of the above wave-functions (or constant coefficients in the case of the brane pseudo-scalars) are vanishing. 
We will require
that the wave-functions in eq.\eqref{KKexp}  form complete sets, and that after  substituting the expansion and performing the integration over the extra-dimensional variable $y$, a \emph{diagonal} bilinear Lagrangian result, \emph{i.e.} the fields defined in eq. \eqref{KKexp} are the mass eigenstates. These requests lead to an equation of motion plus a set of BCs that the profiles must satisfy. Details on the derivation are given in Appendix \ref{appA}; here we only report the results.

Since the proof of the diagonalization is somewhat technical, we will proceed in reverse order, first defining the three sectors of the model, together with the conditions that the corresponding wave-functions have to satisfy, then show how the three sectors are derived by the request of diagonalizing the KK expanded Lagrangian. 
The three sectors are:
\begin{itemize}

\item A \emph{left charged} sector coming from the expansion of the $(W_{L}^{1, \, 2})_M$ fields and the brane pseudo-scalars $\pi^{1,2}$. The explicit form of the expansion is
\begin{equation}
\label{KKexpL}
\begin{split}
& W_{L \; \mu}^{1,2} (x,y) = \sum_{n = 0}^\infty f_{L \, n}^{1,2} (y) \, W_{L \, \mu}^{1,2 \, (n)}(x),\\
& W_{L \; 5}^{1,2} (x,y) = \sum_{n = 0}^\infty g_{L \, n}^{1,2} (y) \, G_L^{1,2 \, (n)}(x),\\
& \pi^{1,2} (x) = \sum_{n = 0}^\infty c_n^{1,2} \, G^{(n)}(x).
\end{split}
\end{equation}
The wave-functions of the vector fields satisfy the equation of motion:
\begin{equation}
\label{KKeomL}
\hat{D} f_{L \, n}^{1,2} = - m_{L \, n}^2 f_{L \, n}^{1,2},
\end{equation}
where we defined the differential operator:
\begin{equation}
\label{hatD}
\hat{D} \equiv \de_y (b(y) \de_y (\cdot)),
\end{equation}
and the set of BCs:
\begin{align}
\label{KKBC0L}
& \quad \de_y f_{L \, n}^{1,2} = 0 \qquad && \text{at } y = 0,\\
\label{KKBCpiRL}
& \left(\tfrac{\tilde{g}^2}{g_5^2} \de_y - b(\pi R) \ m_{L\,n}^2 + \tfrac{\tilde{g}^{2}\tilde{v}^2}{4} \right) f_{L \, n}^{1, \, 2} = 0 && \text{at } y = \pi R.
\intertext{The pseudo-scalar profiles are fixed by the conditions:}
\label{KKscalarsL}
& \quad g_{L \, n}^{1,2} = \frac{1}{m_{L \, n}} \de_y f_{L \, n}^{1,2}, \qquad c_n^{1, \, 2} = \frac{\tilde{v}}{2 m_{L \, n}} f_{L \, n}^{1, \, 2} \big|_{\pi R}.
\end{align}
Note that in this sector no massless solution is allowed; in fact, eq. \eqref{KKeomL} together with the Neumann BC at $y = 0$ \eqref{KKBC0L} imply that a massless mode must have a \emph{constant} profile, and a constant, massless solution cannot satisfy the BC at $y = \pi R$ \eqref{KKBCpiRL}. Also note that eq. \eqref{KKeomL} and the BCs \eqref{KKBC0L} and \eqref{KKBCpiRL} are diagonal in the isospin index, so we have $f_{L \, n}^1 = f_{L \, n}^2$.

Some caution must be used in writing down the completeness and orthogonality relations for the $f_{L \, n}^{1,2}$ mode functions. The differential operator $\hat{D}$   \eqref{hatD} is in fact not hermitian with respect to the ordinary scalar product when evaluated on functions obeying BCs of the kind \eqref{KKBCpiRL}, due to the presence of terms explicitly containing the eigenvalues $m_{L \, n}$ which are induced by $\pi R$-localized terms in the action. To obtain the correct completeness and orthogonality properties of this function set, a generalized scalar product must be used which takes into account such terms. This is given by
\begin{equation}
\label{dotproL}
\left(f_{L \, n}^{1, \, 2}, f_{L \, m}^{1, \, 2}\right)_{\tilde{g}} = L_m^2 \delta_{m n}, \quad (f,h)_{\tilde{g}} = \frac{1}{g_5^2} \int_0^{\pi R} dy \ f h + \frac{1}{\tilde{g}^2} \left. f h \right|_{\pi R},
\end{equation}
where $L_m$ sets the normalization. Since the scalar product $(\cdot,\cdot)_{\tilde{g}}$ is dimensionless, we will set: $L_m \equiv 1$. This will ensure that the kinetic terms of the bosons of this sector are canonically normalized. From this definition we deduce the completeness relation:
\begin{equation}
\label{completeness}
\begin{split}
 \frac{1}{g_5^2} \sum_k f_{L \, k}^{1,\, 2} (y) f_{L \, k}^{1,\, 2} (z) + \frac{1}{\tilde{g}^2} \delta (z-\pi R) \sum_k f_{L \, k}^{1,\, 2} (y) f_{L \, k}^{1,\, 2} (\pi R)= \delta(y-z);
\end{split}
\end{equation}

\item A \emph{right charged} sector coming from the expansion of $(W_{R}^{1, \, 2})_M$. The explicit form of the expansion for this sector is
\begin{equation}
\begin{split}
& W_{R \; \mu}^{1,2} (x,y) = \sum_{n = 0}^\infty f_{R \, n}^{1,2} (y) \, W_{R \, \mu}^{1,2 \, (n)}(x),\\
& W_{R \; 5}^{1,2} (x,y) = \sum_{n = 0}^\infty g_{R \, n}^{1,2} (y) \, G_R^{1,2 \, (n)}(x).
\end{split}
\end{equation}
The wave-functions of the vector fields satisfy a similar equation of motion:
\begin{equation}
\label{KKeomR}
\hat{D} f_{R \, n}^{1,2} = - m_{R \, n}^2 f_{R \, n}^{1,2},
\end{equation}
and the set of BCs:
\begin{align}
\label{KKBC0R}
& \de_y f_{R \, n}^{1,2} = 0 && \qquad \text{at } y = 0,\\
\label{KKBCpiRR}
& f_{R \, n}^{1,2} = 0 && \qquad \text{at } y = \pi R.
\intertext{The scalar profiles are given by:}
\label{KKscalarsR}
&
 \quad g_{R \, n}^{1,2} = \frac{1}{m_{R \, n}} \de_y f_{R \, n}^{1,2}.
\end{align}
Again, in this sector there is no massless solution, for the constant profile of a massless mode is incompatible with the BC \eqref{KKBCpiRR}. Also, the equation of motion and the BCs are again diagonal in the isospin index, so $f_{R \, n}^1 = f_{R \, n}^2$. The right charged sector obeys the usual $L^2$ orthogonality property:
\begin{equation}
\label{dotproR}
\left(f_{R \, n}^{1, \, 2}, f_{R \, m}^{1, \, 2}\right) \equiv \frac{1}{g_5^2} \int_0^{\pi R} dy \ f_{R \, n}^{1, \, 2} f_{R \, m}^{1, \, 2} = R_m^2 \delta_{m n},
\end{equation}
where the factor $1/g_5^2$ has been inserted to compensate for the mass dimension of the integral, so that we can normalize: $R_m \equiv 1$, again ensuring that the kinetic terms will have the canonical normalization.

\item Finally, a \emph{neutral} sector coming from the expansion of $(W_L^3)_M$,  $(W_R^3)_M$ and $\pi^3$. The expansion has the form
\begin{equation}
\begin{split}
& W_{L \; \mu}^3 (x,y) = \sum_{n = 0}^\infty f_{L \, n}^3 (y) \, N_\mu^{(n)}(x), \quad W_{L \; 5}^3 (x,y) = \sum_{n = 0}^\infty g_{L \, n}^3 (y) \, G_N^{(n)}(x),\\
& W_{R \; \mu}^3 (x,y) = \sum_{n = 0}^\infty f_{R \, n}^3 (y) \, N_\mu^{(n)}(x), \quad W_{R \; 5}^3 (x,y) = \sum_{n = 0}^\infty g_{R \, n}^3 (y) \, G_N^{(n)}(x),\\
& \mspace{180mu} \pi^3 (x) = \sum_{j = 0}^\infty c_j^3 \, G^{(j)}(x);
\end{split}
\end{equation}
the equation of motion and the BCs for the vector profiles are given by:
\begin{align}
\label{KKeomN}
& \hat{D} f_{L, R \, n}^3 = - m_{N \, n}^2 f_{L, R \, n}^3,\\
\label{KKBC0N}
& \de_y f_{L, R \, n}^3 = 0 && \text{at } y = 0,\\
\label{KKBCpiRN}
& \mspace{-20mu} \left\{ \begin{aligned}
& \left(\tfrac{\tilde{g}^2}{g_5^2} \de_y - b(\pi R) \ m_{N \, n}^2 + \tfrac{\tilde{g}^{2}\tilde{v}^2}{4} \right) \ f_{L \, n}^3 - \tfrac{\tilde{g}^{2}\tilde{v}^2}{4} \ f_{R \, n}^3 = 0\\
& \left(\tfrac{\tilde{g}^{'2}}{g_5^2} \de_y - b(\pi R) \ m_{N \, n}^2 +\tfrac{\tilde{g}^{'2}\tilde{v}^2}{4} \right) \ f_{R \, n}^3 - \tfrac{\tilde{g}^{'2}\tilde{v}^2}{4} \ f_{L \, n}^3 = 0
\end{aligned} \right. && \text{at } y = \pi R,\\
\intertext{and the pseudo-scalar profiles satisfy}
& g_{L, R \, n}^3 = \frac{1}{m_{N \, n}} f_{L, R \, n}^3, && \mspace{-50mu} \text{if } m_{N \, n} \neq 0 \notag\\
\label{KKscalarsN}
& g_{L, R \, n}^3 = 0 && \mspace{-50mu} \text{if } m_{N \, n} = 0\\
& c_n^3 = \frac{\tilde{v}}{2 m_n} \left. \left(f_{L \, n}^3 - f_{R \, n}^3 \right) \right|_{\pi R}. \notag
\end{align}
In contrast with the charged ones, the neutral sector admits a single massless solution; we have $m_{N \, 0} = 0$. Eqs. \eqref{KKeomN} and \eqref{KKBC0N} imply for a massless mode that both $f_{L \, n}^3$ and $f_{R \, n}^3$ must be constant; then, using also eq. \eqref{KKBCpiRN} we get:
\begin{equation}
\label{charge}
f_{L \, 0}^3 = f_{R \, 0}^3 \equiv f_0,
\end{equation}
where $f_0$ is a constant. The massless mode has to be identified with the photon $\To N_\mu^{(0)} \equiv A_\mu$; since it is the only massless mode in the spectrum we have that the symmetry of the vacuum is, correctly, just $U(1)_{e.m.}$. The ``charged'' and ``neutral'' labels we have given to the three sectors refer to their transformation properties with respect to this unbroken symmetry.

As in the case of the left charged sector, the BCs at $y = \pi R$ in this case explicitly contain the mass of the $n^{th}$ mode, so that again the basis wave-functions $f_{L \, n}^3$ and $f_{R \, n}^3$ have nonstandard orthogonality properties. The correct relations are:
\begin{equation}
\label{normN1}
\left(f_{L \, n}^3, f_{L \, m}^3 \right)_{\tilde{g}} = (N_m^L)^2 \delta_{m n}, \quad \left(f_{R \, n}^3, f_{R \, m}^3 \right)_{\tilde{g}'} = (N_m^R)^2 \delta_{m n},
\end{equation}
where $(\cdot,\cdot)_{\tilde{g}'}$ is defined in a way analogous to $(\cdot,\cdot)_{\tilde{g}}$ (eq. \eqref{dotproL}). Completeness relations similar to that in eq. \eqref{completeness} also hold. Note that it is not possible to set both $N_n^L$ and $N_n^R$ to 1. In fact, since they obey the same differential equation \eqref{KKeomN} and the same BC at $y=0$ \eqref{KKBC0N}, $f_{L \, n}^3$ and $f_{R \, n}^3$ are proportional to each other:
\begin{equation}
\label{Kn}
f_{L \, n}^3 = K_n f_{R \, n}^3,
\end{equation}
and the constants $K_n$ are fixed by the BCs at $y=\pi R$ \eqref{KKBCpiRN}. To get, also in this case,
 canonically normalized kinetic terms we have to set:
\begin{equation}
\label{normN2}
(N_n^L)^2 + (N_m^R)^2 = 1;
\end{equation}
the ratio $(N_n^L)/(N_n^R)$ will be fixed by the value of $K_n$ and by eq. \eqref{normN1}. In particular, for the massless mode it is easy to get
\begin{equation}
\label{masslessfunc}
\frac{1}{f_0^2} = \frac{2 \pi R}{g_5^2} + \frac{1}{\tilde{g}^2} + \frac{1}{\tilde{g}^{'2}}.
\end{equation}

\end{itemize}

\subsection{The expanded Lagrangian}

After the expansion in mass eigenstates, the gauge Lagrangian,
taking into account contributions from both brane and bulk terms,
 is reduced to the form:
\begin{equation}
\label{diagL}
\begin{split}
\lag_{gauge} = & - \frac{1}{2} W_{L \, \mu \nu}^{+ \, (n)} W_{L}^{- \, (n) \; \mu \nu} - \frac{1}{2} W_{R \, \mu \nu}^{+ \, (n)} W_{R}^{- \, (n) \; \mu \nu} - \frac{1}{4} N_{\mu \nu}^{(n)} N^{(n) \; \mu \nu}\\
& - \left|\de_\mu G_L^{+ \, (n)} - m_{L \, n} W_{L \, \mu}^{+ \, (n)} \right|^2 - \left|\de_\mu G_R^{+ \, (n)} - m_{R \, n} W_{L \, \mu}^{+ \, (n)} \right|^2\\
& - \frac{1}{2} \left(\de_\mu G_N^{(n)} - m_{N \, n} N_\mu^{(n)} \right)^2\\
+ \, \bigg\{ i \, g^{L}_{klm} & \left[N_{\mu \nu}^{(m)} W_{L}^{+ \, (k) \; \mu} W_{L}^{- \, (l) \; \nu} + N_\mu^{(m)}(W_L^{- \, (l) \; \mu \nu}) W_{L \, \nu}^{+ \, (k)} - h.c.)\right]\\
+ \ g_{klmn}^{2 \ LL} & \Big[W_L^{+ \, (k) \; \mu} W_L^{- \, (l) \; \nu} W_L^{+ \, (m) \; \rho} W_L^{- \, (n) \; \sigma} (\eta_{\mu \rho} \eta_{\nu \sigma} - \eta_{\mu \nu} \eta_{\rho \sigma}) \Big] \\
& \mspace{-90mu} + \ g_{klmn}^{2 \ LN} \Big[W_L^{+ \, (k) \; \mu} W_L^{- \, (l) \; \nu} N^{(m) \; \rho} N^{(n) \; \sigma} (\eta_{\mu \rho} \eta_{\nu \sigma} - \eta_{\mu \nu} \eta_{\rho \sigma}) \Big] + (L \leftrightarrow R) \bigg\}\,.
\end{split}
\end{equation}
 The bilinear part of the Lagrangian is, as announced, diagonal. The trilinear and quadrilinear coupling constants $g^{L}_{klm}, \ g_{klmn}^{2 \ LL}, \ g_{klmn}^{2 \ LN}$ are defined in terms of the gauge profiles:
\begin{align}
\label{tricoupl}
g_{klm}^L & = \frac{1}{g_5^2} \int_0^{\pi R} dy f_{L \, k}^1 f_{L \, l}^1 f_{L \, m}^3 + \frac{1}{\tilde{g}^2} f_{L \, k}^1 f_{L \, l}^1 f_{L \, m}^3 |_{\pi R},\\
g_{klmn}^{2 \ LL} & = \frac{1}{g_5^2} \int_0^{\pi R} dy  f_{L \, k}^1 f_{L \, l}^1 f_{L \, m}^1 f_{L \, n}^1 + \frac{1}{\tilde{g}^2}  f_{L \, k}^1 f_{L \, l}^1 f_{L \, m}^1 f_{L \, n}^1 |_{\pi R},\\
g_{klmn}^{2 \ LN} & = \frac{1}{g_5^2} \int_0^{\pi R} dy  f_{L \, k}^1 f_{L \, l}^1 f_{L \, m}^3 f_{L \, n}^3 + \frac{1}{\tilde{g}^2}  f_{L \, k}^1 f_{L \, l}^1 f_{L \, m}^3 f_{L \, n}^3 |_{\pi R}
\end{align}
(remember that $f_{L(R) \, n}^1 \equiv f_{L(R) \, n}^2$); similar definitions hold for the coupling constants $g^{R}_{klm}, \ g_{klmn}^{2 \ RR}, \ g_{klmn}^{2 \ RN}$ of the right sector, but without any contribution from boundary terms due to eq. \eqref{KKBCpiRR}.

An important observation  concerns the couplings $g_{kl0}^{L(R)}$. These give the coupling of $N^{(0)}$ , which we identified with the photon, with the charged fields; as a consequence, they should all be equal to the electric charge, for any value of $k, \ l$. By the definition \eqref{tricoupl} and eq. \eqref{charge}, we immediately get:
\begin{equation}
g_{kl0}^L = g_{kl 0}^R \equiv f_0 \delta_{kl},
\end{equation}
thanks to the fact that the wave-functions $f_{L \, k}^1$ and $f_{R \, k}^1$ form an orthonormal basis. Then we conclude that
\begin{equation}
f_0 = e.
\end{equation}
Then, from Eq.~\eqref{masslessfunc}, we derive an expression for the electric charge as a function of the model parameters:
\begin{equation}
\label{charge2}
\frac{1}{e^2} = \frac{2 \pi R}{g_5^2} + \frac{1}{\tilde{g}^2} + \frac{1}{\tilde{g}^{'2}}.
\end{equation}

The actual profiles and masses can of course only be obtained by specifying the warp factor $b(y)$. However, it is possible to write, in general, the  equations from \eqref{KKexpL} to \eqref{KKscalarsN}) in a more compact form. In fact, equations of motion \eqref{KKeomL}, \eqref{KKeomR} and \eqref{KKeomN} all have the same  form, $\hat{D} f = -m^2 f$. This is a second order ODE, so it admits two independent solutions. Following ref. \cite{Falkowski:2007iv}, we can introduce two convenient linear combinations $C(y,m_n)$ and $S(y,m_n)$ (``warped sine and cosine'') such that
\begin{equation}
\label{wCS}
C(0,m) = 1, \quad \de_y C(0,m)=0; \qquad S(0,m) = 0, \quad \de_y S(0,m) = m
\end{equation}
with $m \neq 0$ (we have already seen that there is a single massless mode and that its profile is constant). In the limit of a flat extra dimension, these functions reduce to the ordinary sine and cosine.

Thanks to the Neumann BCs on the $y = 0$ brane \eqref{KKBC0L}, \eqref{KKBC0R}, \eqref{KKBC0N}, the vector profiles $f_{L, R \, n}^a$ are all proportional to $C(y,m_n)$. The eigenvalues, that is the physical masses of the vector fields $m_{L \, n}$, $m_{R \, n}$ and $m_{N \, n}$, are then fixed by the BCs on the IR brane \eqref{KKBCpiRL}, \eqref{KKBCpiRR} and \eqref{KKBCpiRN}. For the three sectors we can easily derive three eigenvalue equations:
\begin{align}
\intertext{\emph{Left charged:}}
\label{leftmodes}
& \tfrac{\tilde{g}^2}{g_5^2} C'(\pi R,m_{L \, n}) - (b(\pi R) \ m_{L \, n}^2 - \tfrac{\tilde{g}^2\tilde{v}^2}{4}) \, C(\pi R,m_{L \, n}) = 0 \\
\intertext{\emph{Right charged:}}
\label{rightmodes}
& C(\pi R,m_{R \, n}) = 0 \\
\intertext{\emph{Neutral:}}
\label{neutmodes}
\begin{split}
& \Big(\tfrac{\tilde{g}^2}{g_5^2} C'(\pi R,m_{N \, n}) - (b(\pi R) \ m_{N \, n}^2 - \tfrac{\tilde{g}^2\tilde{v}^2}{4}) \, C(\pi R,m_{N \, n}) \Big) \cdot \\
& \Big(\tfrac{\tilde{g}^{'2}}{g_5^2} C'(\pi R,m_{N \, n})  - (b(\pi R) \ m_{N \, n}^2 - \tfrac{\tilde{g}^{'2}\tilde{v}^2}{4}) \, C(\pi R,m_{N \, n}) \Big)\\
& \ = \tfrac{\tilde{g}^2 \tilde{g}^{'2} \tilde{v}^4}{16} C(\pi R,m_{N \, n})^2
\end{split}
\end{align}
In section \ref{phen} we will make extensive use of these equations for specific choices of the warp factor and of the parameters of the models to obtain explicit examples of the KK spectrum.

\section{Low energy limit and EW precision observables}
\label{lowen5D}

We can obtain a convenient low-energy approximation of the theory by using the so-called \emph{holographic} approach \cite{Witten:1998qj,ArkaniHamed:2000ds,Rattazzi:2000hs,PerezVictoria:2001pa,Barbieri:2003pr,Burdman:2003ya,Luty:2003vm}, which consists in integrating out the bulk degrees of freedom in the functional integral. For the purposes of the present calculation it is sufficient to take into account just the tree-level effects of the heavy resonances, so the integration can be done by simply eliminating the bulk fields from the Lagrangian via their classical equations of motion; moreover, bulk gauge self-interactions can be neglected.

The equations to be solved are:
\begin{equation}
\hat{D} \, W_{L(R) \; \mu}^a (p,y) = (p^2\delta_{\mu \nu} - p_\mu p_\nu) 
W_{L(R)}^{a \; \nu} (p,y)
\end{equation}
with $\hat D$ defined in eq. \eqref{hatD}) and we have Fourier transformed with respect to the first four coordinates. As previously discussed, on the $y = 0$ brane, we do not want to make any assumptions on the value of the fields; so we leave their variations arbitrary, and since there are no localized terms on the brane, this leads to Neumann boundary conditions for all of the fields:
\begin{equation}
\label{holoBC1}
\left\{ \begin{aligned}
         & \de_y W_{L \; \mu}^a = 0\\
         & \de_y W_{R \; \mu}^a = 0
         \end{aligned} \right. \qquad y = 0.
\end{equation}
At the other end of the AdS segment, we account for the presence of localized terms by imposing four fields to be equal to generic \emph{source fields}, while the other two (those corresponding to the right charged sector) are vanishing:
\begin{equation}
\label{holoBC2}
\left\{ \begin{aligned}
         &  W_{L \; \mu}^a = \tilde{W}_\mu^a\\
         &  W_{R \; \mu}^3 = \tilde{B}_\mu\\
         & W_{R \; \mu}^{1,2} = 0
         \end{aligned} \right. \qquad y = \pi R.
\end{equation}
The first step in solving the equations is to split the fields in their \emph{longitudinal} (aligned with $p_\mu$) and \emph{transverse} parts. The operator $(p^2\delta_{\mu \nu} - p_\mu p_\nu)$ is vanishing when acting on the longitudinal part, while it is simply equivalent to $p^2$ when acting on the transverse one. In this way, each equation can be split into two simpler ones:
\begin{equation}
\label{holoeq2}
\left\{ \begin{aligned}
         & \hat{D} \, W_{L/R \; \mu}^{a, \; tr} = p^2 \, W_{L/R \; \mu}^{a, \; tr}\\
         & \hat{D} \, W_{L/R \; \mu}^{a, \; long} = 0
         \end{aligned} \right.
\end{equation}
Taking into account the boundary conditions \eqref{holoBC1}, \eqref{holoBC2}, and defining $|p^2| \equiv \omega^2$, eqs. \eqref{holoeq2} are simply solved; the solutions are given by
\begin{equation}
\left\{ \begin{aligned}
         & W_{L \; \mu}^{a, \; tr} = (\tilde{C}(y,\omega) - \frac{\tilde{C}'(0,\omega)}{\tilde{S}'(0,\omega)} \tilde{S}(y,\omega)) \tilde{W}_\mu^{a, \; tr}\\
         & W_{L \; \mu}^{a, \; long} = \tilde{W}_\mu^{a, \; long}\\
         & W_{R \; \mu}^{3, \; tr} = (\tilde{C}(y,\omega) - \frac{\tilde{C}'(0,\omega)}{\tilde{S}'(0,\omega)} \tilde{S}(y,\omega)) \tilde{B}_\mu^{tr}\\
         & W_{R \; \mu}^{3, \; long} = \tilde{B}_\mu^{long}\\
         & W_{R \; \mu}^{1,2} = 0
         \end{aligned} \right. 
\end{equation}
with
\begin{equation}
\label{SCalt}
\begin{split}
\hat{D} \, (\tilde{S}, \tilde{C}) = & - \omega^2 (\tilde{S},\tilde{C});\\
\tilde{S}(\pi R,\omega) = 0, \ & \tilde{S}'(\pi R,\omega) = \omega;\\
\tilde{C}(\pi R,\omega) = 1, \ & \tilde{C}'(\pi R,\omega)= 0.
\end{split}
\end{equation}
As it can be seen, the first two components of the right sector drop out from the low-energy effective Lagrangian altogether; this corresponds to the fact that in general the right charged sector  does not contain any  light mode, in contrast with the left charged and neutral ones.

Before substituting the solutions, note that the bulk contribution to the Lagrangian in Eq.~\eqref{5Daction}
 can be reduced - through an integration by parts - to a surface term plus a term proportional to the equations of motion,
\begin{equation}
\begin{split}
\mspace{-10mu}\lag_{bulk}^{(2)} = -\frac{1}{2g_5^2} 
& \left(\de_y( b(y) W_{L \; \mu}^{a, \; tr} W_L^{a, \; tr \; \mu}) - W_{L \; \mu}^{a, \; tr} ((\hat{D} - p^2)\delta_\nu^\mu + p^\mu p_\nu) W_L^{a, \; tr \; \nu} \right)\\
& \mspace{100mu} + (L \to R).
\end{split}
\end{equation}
After the substitution, most of the terms vanish due to the BCs; we are  left with
\begin{equation}
\label{presol}
\lag_{bulk}^{(2)} = - \frac{1}{2g_5^2} b(y) W_{L \; \mu}^a  \ \de_y W_L^{a \; \mu}|_{\pi R} + (L \to R, \ a \to 3);
\end{equation}
taking into account the definition of the $\tilde{C}$ and $\tilde{S}$ functions \eqref{SCalt}, eq. \eqref{presol} reduces to
\begin{equation}
\label{presol2}
\lag_{bulk}^{(2)} = \frac{\omega \ b(\pi R)}{2g_5^2}  \left.\frac{\tilde{C}'}{\tilde{S}'}\right|_0 \left(\tilde{W}_\mu^{a, \; tr} \ \tilde{W}^{a, \; tr \; \mu} + \tilde{B}_\mu^{tr} \ \tilde{B}^{tr \; \mu} \right).
\end{equation}
Eq. \eqref{presol2} has a complicated dependence on $\omega$ hidden in the functions $\tilde{S}'|_0$, $\tilde{C}'|_0$. In order to extract the low-energy behaviour of the theory, let's expand in $\omega$. This can be done in general, without needing to specify $b(y)$. In fact, using eq. \eqref{SCalt}, it is not difficult to show that the functions $\tilde{C}(y,\omega)$ and $\tilde{S}(y,\omega)$ obey the integral equations:
\begin{align}
& \tilde{C}(y,\omega) = 1 - \omega^2 \int_y^{\pi R} dy' \, b^{-1}(y')\int_{y'}^{\pi R} dy'' \, \tilde{C}(y'',\omega)\\
& \mspace{-50mu} \tilde{S}(y,\omega) = \omega \int_y^{\pi R} dy' \, b^{-1}(y')  - \omega^2 \int_y^{\pi R} dy' \, b^{-1}(y')\int_{y'}^{\pi R} dy'' \, \tilde{S}(y'',\omega),
\intertext{from which we can derive a low-energy expansion (small $\omega$):}
\tilde{C}(y,\omega) & = 1 - \omega^2 \int_y^{\pi R} dy' \, y' \, b^{-1}(y') \notag\\
\label{lowC}
& + \omega^4 \int_y^{\pi R} dy' \, b^{-1}(y') \int_{y'}^{\pi R} dy'' \, \int_{y''}^{\pi R} dy''' \, y'''  b^{-1}(y''') + \dots\\
\tilde{S}(y,\omega) & = \omega \int_y^{\pi R} dy' \, b^{-1}(y') \notag\\
\label{lowS}
& - \omega^3 \int_y^{\pi R} dy' \, b^{-1}(y') \int_{y'}^{\pi R} dy'' \, \int_{y''}^{\pi R} dy''' \, b^{-1}(y''') + \dots
\end{align}

We will substitute expansions \eqref{lowC}, \eqref{lowS} in eq. \eqref{presol2}, keeping terms up to $O(\omega^4)$, which - as we will soon show - will reproduce the Standard model  plus corrections of order $m_Z^2/\bar{M}^2$, where $\bar{M}$ is given by 
\footnotemark
\footnotetext{
Notice that the parameter $\bar M$ can be related to the integrals introduced
in \cite{Delgado:2007ne},
\begin{equation}
\frac{1}{\bar{M}^2} = I_2(\pi R) - I_1(\pi R)\nonumber
\end{equation}
where
\begin{equation}
I_{1}(y)=\frac 1 {\pi R}\int_0^y\int_0^z dz' \, z' b^{-1}(z'), \quad I_{2}(y) = \int_0^y dz' \, z' b^{-1}(z')\nonumber
\end{equation}}
:
\begin{equation}
\label{doubleint}
\frac{1}{\bar{M}^2} = \frac{1}{\pi R} \int_0^{\pi R} dy \int_y^{\pi R} dz \; z \; b^{-1}(z)
\end{equation}
and, as we will show later, is of the order of the mass of the lightest resonance that we have integrated out. After the substitution, the bulk Lagrangian becomes
\begin{equation}
\label{sol}
\begin{split}
\lag_{bulk}^{(2)} = & - \frac{\pi R}{2 g_5^2} \Big(\tilde{W}_\mu^a (p^2 \eta^{\mu \nu} - p^\mu p^\nu)(1 - \frac{p^2}{\bar{M}^2}) \tilde{W}_\nu^a\\
& + \tilde{B}_\mu (p^2 \eta^{\mu \nu} - p^\mu p^\nu)(1 - \frac{p^2}{\bar{M}^2}) \tilde{B}_\nu \Big);
\end{split}
\end{equation}

Finally, let us add the contribution  coming from the brane. Switching back to the  coordinate space, the final expression is
\begin{equation}
\label{leff5D}
\begin{split}
\lag_{eff} = & - \frac{1}{4g^2} \tilde{W}_{\mu \nu}^a \tilde{W}^{a \; \mu \nu} - \frac{1}{4g^{'2}} \tilde{B}_{\mu \nu} \tilde{B}^{\mu \nu}\\
& -  \frac{v^2}{8} \left(\tilde{W}_\mu^a \tilde{W}^{a \; \mu} + \tilde{B}_\mu \tilde{B}^\mu - 2 \tilde{W}_\mu^3 \tilde{B}^\mu \right)\\
& + \frac{\pi R}{4 g_5^2} \left (\tilde{W}_{\mu \nu}^a \frac{\square}{\bar{M}^2} \tilde{W}^{a \; \mu \nu} +  \tilde{B}_{\mu \nu} \frac{\square}{\bar{M}^2} \tilde{B}^{\mu \nu}\right )\\
+ & \ \textrm{\textit{bosonic self-interactions}} + \textrm{\textit{fermion terms}}
\end{split}
\end{equation}
where we have introduced the effective couplings
\begin{equation}
\label{effcoupl}
\frac{1}{g^2} = \frac{1}{\tilde{g}^2} + \frac{1}{{\bar g}_5^2}, \qquad \frac{1}{{g'}^{2}} = \frac{1}{\tilde{g}^{'2}} + \frac{1}{{\bar g}_5^2}.
\end{equation}
and
\begin{equation}
 v = \tilde{v} \, b(\pi R), \quad {\bar g}_5^2=g_5^2/\pi R.
\end{equation}

The same result was obtained in the deconstructed GD-BESS model  \cite{Casalbuoni:2007dk}. In fact from the correspondence in Eq.~ \eqref{identification}, taking   the gauge couplings $g_i=g_c$, we get:

\begin{equation}
  \frac {1}{{\bar g}_5^2}\rightarrow \frac 1 {{\overline G}^2}=\sum_{k=1}^N
\frac 1{g_k^2}=\frac N{g_c^2}
\label{Gbarra}
\end{equation}

\begin{equation}
\left. \frac{1}{\bar{M}^2} \right|_{decon.} = \sum_{i=1}^N \frac{1}{g_c^2} \sum_{j = N+i}^{2N+1} \frac{j-N}{f_j^2 g_c^2} = \sum_{i=1}^N \frac{1}{g_c^2} \sum_{j = i}^{N+1} \frac{j}{f_j^2 g_c^2},
\label{mbarra}
\end{equation}
The Eq.~\eqref{mbarra} in the continuum limit reproduces
Eq.~\eqref{doubleint}.

Starting from Eq.~\eqref{leff5D}, we can mirror the calculation done in ref. \cite{Casalbuoni:2007dk} to obtain the seven form factors encoding the corrections from new physics to the EW precision observables \cite{Barbieri:2004qk},
and  then the
$\epsilon$ parameters. 
The results are 
\begin{equation}
\hat S=\hat T=\hat U=V=X=0
\end{equation}
\begin{equation}
W=\frac {g^2 c_\theta^2 m_Z^2}{\bar{M}^2\bar{g}_5^2},\,\,\,
Y=\frac {g^{'2} c_\theta^2 m_Z^2}{\bar{M}^2\bar{g}_5^2}
\end{equation}
with $\tan\theta = g'/g$  and
\begin{align}
& \epsilon_1 = - \frac{(c_{\theta}^4 + s_{\theta}^4)}{c_{\theta}^2} \overline{X}, \quad \epsilon_2 = - c_{\theta}^2 \overline{X}, \quad \epsilon_3 = - \overline{X}
\end{align}
with
\begin{equation}
\overline{X} = \frac{m_Z^2}{\bar{M}^2} \left(\frac{g}{\bar{g}_5}\right)^2.
\end{equation}

Note that $\epsilon_{1,2,3}$ are all proportional to $\overline X$ which contains a double suppression factor. This feature was the main ingredient for the compatibility of
the D-BESS model with the EW precision tests. In the five dimensional formulation of this model the ratio $(g/\bar g_5)^2$ originates from the presence of brane localized
kinetic terms.

The $\epsilon$ parameters can be tested against the experimental data.
 To do this, 
we need to express the model parameters in terms of the physical quantities.
Proceeding again as in \cite{Casalbuoni:2007dk} we get the expressions the standard input parameters
 $\alpha$, $G_F$ and $m_Z$ in terms of the model parameters. For convenience we rewrite the results:
\begin{align}
\label{alpha5D}
& \alpha \equiv \frac{e^2}{4 \pi} = \frac{g^2 s_\theta^2}{4 \pi},\\
\label{emmez}
& m_{Z}^2 = \tilde{M}_Z^2 \left(1 - z_Z \frac{\tilde{M}_Z^2}{\bar{M}^2} \right),\,\,\, m_{W}^2 = \tilde{M}_W^2 \left(1 - z_W \frac{\tilde{M}_W^2}{\bar{M}^2} \right),\\
\label{GF5D}
& \frac{G_F}{\sqrt{2}} \equiv \frac{e^2}{8s_{\theta_0}^2 c_{\theta_0}^2 m_Z^2}, \quad  \quad s_{\theta_0}^2 c_{\theta_0}^2 = s_{\theta}^2 c_{\theta}^2 \left(1 + z_Z \frac{m_Z^2}{\bar{M}^2} \right),
\end{align}
with 
\begin{equation}
  z_Z = \frac{g^2 (c_{\theta}^4 + s_{\theta}^4)}{c_{\theta}^2 \, \bar{g}_5^2},\,\,  \quad \tilde{M}_Z^2 = \frac{v^2 (g^2 + {g'}^{2})}{4}
\end{equation}
\begin{equation}
  z_W = \frac{g^2 }{ \bar{g}_5^2},\,\,  \quad \tilde{M}_W^2 = \frac{v^2 g^2 }{4}
\end{equation}

In section \ref{phen}, we will study the constraints on the model parameter space by EW precision parameter for two choices of the warp factor, $b(y) \equiv 1$ (flat extra dimension) and $b(y) = e^{-2k y}$ (a slice of AdS$_5$). To this aim
we need to invert 
  \eqref{alpha5D}, \eqref{emmez} and \eqref{GF5D},
\begin{align}
\label{gpar}
& g^2 = \frac{4\pi \alpha}{s_{\theta_0}^2} \left(1 + \frac{4\pi 
\alpha(c_{\theta_0}^4 + s_{\theta_0}^4)}{\bar{g}_5^2 s_{\theta_0}^2 
c_{2\theta_0}} \frac{m_Z^2}{\bar{M}^2}\right),\\
\label{g1par}
& g^{'2} = \frac{4\pi \alpha}{c_{\theta_0}^2} \left(1 - \frac{4\pi 
\alpha(c_{\theta_0}^4 + s_{\theta_0}^4)}{\bar{g}_5^2 c_{\theta_0}^2 c_{2\theta_0}} \frac{m_Z^2}{\bar{M}^2}\right),\\
\label{vpar}
& {v}^2 = \frac{4 }{g^2+g^{'2}} m_Z^2 \left(1+ \frac{4\pi \alpha (c_{\theta_0}^4 + s_{\theta_0}^4)}{\bar{g}_5^2 s_{\theta_0}^2  c_{\theta_0}^2} \frac{m_Z^2}{\bar{M}^2}\right)=\frac 1 {\sqrt{2} G_F};
\end{align}
then, using definitions \eqref{effcoupl},  obtain also $\tilde{g}$ and $\tilde{g}'$.

\subsection{Notes on unitarity and the Higgs field}
\label{unitarity}

As any gauge theory in 5 space-time dimensions, the 5D-DBESS model has couplings with negative mass dimension and is therefore not renormalizable. In the KK expanded 4D theory emerging from the compactification of the extra dimension, the nonrenormalizability manifests as a partial wave unitarity violation at tree level at an energy scale proportional to the inverse square of the gauge coupling \cite{SekharChivukula:2001hz}. A detailed study of the unitarity properties of the model was beyond the scope of the present work, but it is still possible  to give an estimate based on naive dimensional analysis. In flat space, the naive estimate for a gauge theory with dimensional coupling constant $g_5$ gives a cut-off $\Lambda = (16 \pi^2)/g_5^2$ \cite{Becciolini:2009fu}.

In a warped space, the cut-off is dependent on the location along the fifth dimension: starting from $\Lambda$ at the $y=0$ brane, it is redshifted along the interval (as any other energy scale in the theory), getting down to $\Lambda' = \Lambda \, \sqrt{b(\pi R)}$ upon reaching the $y=\pi R$ brane. To get an estimate for the Kaluza-Klein 4D effective theory, we will use the most restrictive cut-off:
\begin{equation}
\label{unitcutoff}
\Lambda' = \frac{16 \pi^2}{g_5^2} \sqrt{b(\pi R)}.
\end{equation}

In addition to the one coming from the negative mass dimension bulk coupling $g_5$ (or equivalently from the infinite tower of KK excitations), the 5D-DBESS has another, more stringent unitarity bound: 
the one coming from the $WW$ scattering. In this model, in fact, the longitudinal components of the electroweak gauge bosons are only coupled to the $U$ field and,  as a consequence, the corresponding scattering amplitudes violate partial wave unitarity at the same energy scale as in the Higgsless SM \cite{Appelquist:1980vg}, that is $\Lambda_{cut-off} \simeq 1.7$ TeV. The violation of unitarity is not postponed to higher scales as in the 5D Higgsless model \cite{Csaki:2003dt,Csaki:2003zu}. This situation exactly mirrors the one of the GD-BESS model \cite{Casalbuoni:2004id,Casalbuoni:2007dk}.

However, this problem can be easily cured by generalizing the $U$ field to a matrix containing an additional real scalar excitation $\rho$, mimicking  the 
standard Higgs sector in the matrix formulation:
\begin{equation}
U \to M \equiv \frac{\rho}{\sqrt{2}} U.
\end{equation}
Just as in the case of the SM, the exchange of the new scalar degree of freedom $\rho$ cancels the growing with energy terms in the scattering of the longitudinal EW gauge bosons, delaying unitarity violation. A similar process of unitarization via the addition of scalar fields was also studied in the context of the D-BESS model in ref. \cite{Casalbuoni:1997rs}.

We also add self interactions of  the  extra scalar field $\rho$, described by the potential
\begin{equation}
V(\rho) = - \frac{\mu^2}{2} \rho^2 + \frac{\lambda}{4} \rho^4,
\end{equation}
whereupon the field $\rho$ acquires a VEV $\tilde{v} = \frac{\mu}{\sqrt{\lambda}}$ and a mass $m_h = \sqrt{2 b(\pi R)} \mu$. We then expand as usual:
\begin{equation}
\rho = h + \tilde{v};
\end{equation}
in this way the Lagrangian is equal to that of eq. \eqref{5Daction} plus kinetic, mass and interaction terms for $h$. The interactions between $h$ and the gauge bosons help unitarizing the scattering of the longitudinally polarized vectors, and  the unitarity violation is postponed to the scale typical of a 5D theory, $\Lambda'$.

In the GD-BESS case, the presence of a physical scalar was undesirable since it seemed to reintroduce the hierarchy problem. In the continuum limit, however, at least for a particular choice of the extra-dimensional background, the slice of AdS$_5$ that we will analyze in section \ref{RSlimit}, the 
$h$ field can be interpreted as a composite Higgs state - just as the KK excitations of the gauge bosons - by the AdS/CFT correspondence \cite{Maldacena:1997re,Witten:1998qj,ArkaniHamed:2000ds,Rattazzi:2000hs}, sidestepping the hierarchy problem.

\section{Phenomenology}
\label{phen}

In this last section, we are going to do a brief phenomenological study of the continuum GD-BESS in correspondence of two particular choices for the warp factor $b(y)$: the \emph{flat limit}, $b(y) \equiv 1$ and the \emph{RS limit}, $b(y) = e^{-2k y}$. In both cases, we will report spectrum examples, bounds from electroweak precision tests and naive unitarity cut-off.

\subsection{Flat extra dimension}

In  this case, we have $b(y) \equiv 1$. This immediately implies (using eq. \eqref{doubleint})
\begin{equation}
\label{barMflat}
\bar{M} = \frac{\sqrt{3}}{\pi R}.
\end{equation}
To get an interesting phenomenology at an accessible scale, we need $\bar{M} \sim$ TeV. The basic parameters of the model are $\pi R$, the gauge couplings $g_5$, $\tilde{g}$ and $\tilde{g}'$, the VEV of the scalar field $\tilde{v}$ (which is $\equiv v$ since $b=1$) and its self-coupling constant $\lambda$. The latter is only used in the determination of the Higgs mass $m_h$; three out of four of the remaining parameters can be expressed in terms of the three measured quantities that are customarily chosen as input parameters for the SM, $\alpha$, $G_F$ and $m_Z$ using Eqs.~\eqref{gpar},\eqref{g1par},\eqref{vpar}.
 The free parameters of the model are then just $\pi R$ and $g_5$. The order of magnitude of $\pi R$ is fixed by eq. \eqref{barMflat} together with the request $\bar{M} \sim$ TeV, while $\bar{g}_5$ is constrained by eq. \eqref{effcoupl}. In fact, since we need $\tilde{g}^2$ and $\tilde{g}^{'2}$ to be positive, eq. \eqref{effcoupl} implies $\bar{g}_5 > g, \ g'$.

We are now ready to calculate the spectrum. In the flat limit, the $C$ and $S$ functions (eq. \eqref{wCS}) reduce to ordinary trigonometric functions:
$
C(y,m) = \cos(my),\, S(y,m) = \sin(my).
$
However, even in this very simple case only the eigenvalue equation for the right charged sector \eqref{rightmodes} can be analytically solved. We get
\begin{equation}
\label{Rflat}
m_{R \, n} = \frac{2n-1}{2 R}, \quad n = 1, 2, \ldots
\end{equation}
The equations \eqref{leftmodes}, \eqref{neutmodes}, defining the eigenvalues for the other two sectors, have to be solved numerically. Some general remarks can be made at a qualitative level, however.

Eq. \eqref{leftmodes} can be recast in the form
\begin{equation}
\label{leftmodes2}
m_{L \, n} \tan(m_{L \, n} \pi R) = - \frac{g_5^2}{\tilde{g}^2}(m_{L \, n}^2 - \frac{\tilde{g}^2 \tilde{v}^2}{4});
\end{equation}
the eigenvalues of the left charged sector are then determined by the intersection of two curves: the trigonometric curve $\tan(m \pi R)$ and the parabola $- \frac{g_5^2}{\tilde{g}^2}(m^2 - \frac{\tilde{g}^2 \tilde{v}^2}{4})$. The $- \frac{\tilde{g}^2 \tilde{v}^2}{4}$ term - originating from the $y=\pi R$ brane mass term in the action \eqref{5Daction} - raises the vertex of the parabola, allowing for an intersection of the curves near $m=0$. The corresponding  light eigenvalue $m_{L \, 0}$ can be identified with $m_W$. For bigger values of $m$, the parabola goes down as $- m^2$, and the intersections are nearer and nearer the asymptotes of $\tan(m \pi R)$ (which correspond to the zeroes of $\cos(m \pi R)$, and thus to the eigenvalues of the right charged sector, Eq.~\eqref{Rflat}), that are evenly spaced by  $1/R$. The situation is illustrated in Fig.~ \ref{intL}.
\begin{center}
\begin{overpic}[width=0.8\textwidth]{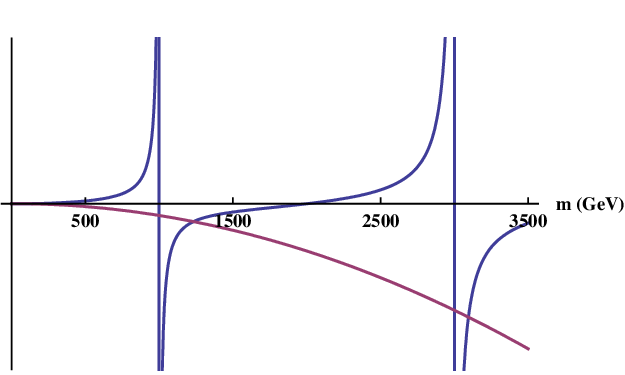}
\put(1,26.7){\framebox[0.048\textwidth]{\rule{0pt}{9pt}}}
\put(75,37){\framebox{\includegraphics[width=0.20\textwidth]{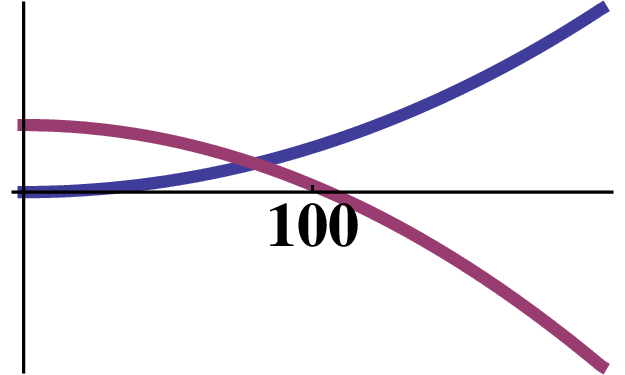}}}
\end{overpic}
\end{center}
\vspace{-0.7cm}
\begin{figure}[h!]
\caption{\small{\textit{ Graphical solutions  of the eigenvalue equation \eqref{leftmodes2} for the left charged sector. The intersections corresponding to the first and second KK excitations can be clearly seen. On the top right panel, 
the magnification of the low mass region showing  the intersection corresponding to the zero mode, W.}}}
\label{intL}
\end{figure}

The neutral sector has a more complicated eigenvalue equation \eqref{neutmodes}. However, it can be easily checked that, as soon as $m \gg m_Z$, the right-hand side of the equation is negligible so that it can be approximated:
\begin{equation}
\begin{split}
& \Big(\tfrac{\tilde{g}^2}{g_5^2} m_{N\, n} \sin(\pi R m_{N \, n}) + (m_{N \, n}^2 - \tfrac{\tilde{g}^2\tilde{v}^2}{4}) \, \cos(\pi R m_{N \, n}) \Big) \cdot \\
& \Big(\tfrac{\tilde{g}^{'2}}{g_5^2} m_{N\, n} \sin(\pi R m_{N \, n})  + (m_{N \, n}^2 - \tfrac{\tilde{g}^{'2}\tilde{v}^2}{4}) \, \cos(\pi R m_{N \, n}) \Big) = 0.\\
\end{split}
\end{equation}
The eigenvalue equation can then be approximately factorized into two independent ones; the first one is identical to eq. \eqref{leftmodes2}, the second one is similar with the replacement $g \to g'$. The tower of the neutral eigenstates is then composed by two sub-towers, one of which almost identical to the one of the left sector. 
In Table \ref{spectrum1table}, we show the lightest part of the spectrum in an explicit example corresponding to a particular choice of the parameters.


\begin{center}
\begin{table}
\begin{tabular}{|c|c|c|c|}
\hline
& \textbf{0 mode  (GeV)} & \textbf{$1^{st}$ KK exc.  (GeV)} & \textbf{$2^{nd}$ KK exc.  (GeV)}\\
\hline
\textit{Left charg.} & 80 & 1232 & 3096 \\
\hline
\textit{Right charg.} & - & 1000 & 3000 \\
\hline
\textit{Neutral} & $0,91$ & $1056,1232$ & $3019,3096$ \\
\hline
\end{tabular}
\caption{\small{\textit{Low lying masses  of the spectrum (zero modes and first two KK excitations) for the model in the flat limit, with the following parameter choice: $\pi R = 1.57 \cdot 10^{-3} \emph{ GeV}^{-1}$, $\bar{g}_5 = 1$,
 naive unitarity cut-off equal to $10^{5}$ \emph{GeV}.}}}
\label{spectrum1table}
\end{table}
\end{center}

\begin{figure}
\includegraphics[width=0.45\textwidth]{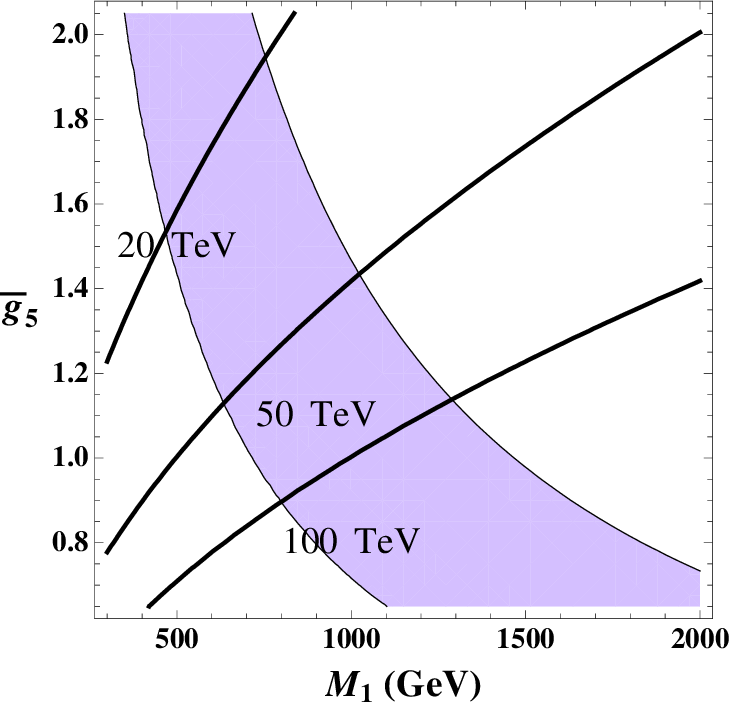}
\includegraphics[width=0.45\textwidth]{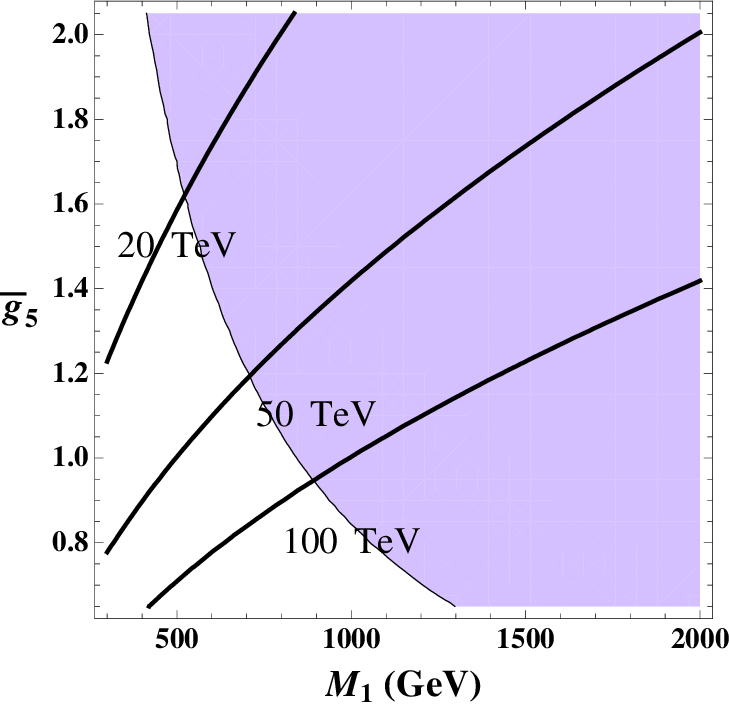}
\caption{\small{\textit{Allowed regions in the $(M_1,\bar{g}_5)$ parameter space for a flat extra dimension, for two values of the Higgs mass: $m_H = 1$ \emph{TeV} (on the left) and $m_H = 300$ \emph{GeV} (on the right), based on electroweak precision constraints. 
 The corresponding naive unitarity cut-off is $10^{5}$ \emph{GeV}, so every shown state is well within the unitarity limit.
Also shown are the  constraints from naive dimensional analysis (contours correspond to different choices of the  UV cut-off).}}}
\label{M1fitb}
\end{figure}

Let us now check  the model against EW precision tests. In Fig.~\ref{M1fitb}, we show the allowed region at $95 \%$ C.L. in parameter space $(M_1,\bar{g}_5)$, based on the new physics contribution to the $\epsilon$ parameters. Here $M_1 \equiv m_{R \, 1} = 1/2R$ is the mass of the lightest KK excitation, that is of the first eigenstate of the right charged sector. The contour is obtained by a $\chi^2$ analysis, based on the following experimental values for the $\epsilon$ parameters:
\begin{align}
& \left.\begin{array}{l}\epsilon_1 = (+5.4 \pm 1.0) 10^{-3} \\ \epsilon_2 = (-8.9 \pm 1.2) 10^{-3}\\ \epsilon_3 = (+5.34 \pm 0.94) 10^{-3} \end{array}\right.\\
\intertext{with correlation matrix}
& \left(\begin{array}{ccc}1 & 0.60 & 0.86 \\ 0.60 & 1 & 0.40 \\0.86 & 0.40 & 1\end{array}\right)\,,
\end{align}
taken from \cite{:2005ema}, and adding to the present model contribution the one from radiative corrections in the SM. To fix the SM contribution, we set  $m_t = 173.1$ \cite{:2009ec}, $\alpha^{-1}(m_Z^2)=128.957\pm 0.020$ \cite{Jegerlehner:2008rs} and consider two different test values of the Higgs mass, $m_H = 1$ TeV and $m_H = 300$ GeV. 
Notice that, since  we have considered SM fermion couplings, the new physics
contribution to $\epsilon_b$ is zero. Since the $\epsilon_b$ experimental
value is very slightly correlated to $\epsilon_{1,2,3}$, we did not include this
observable in the analysis.
We get:
\begin{align}
& \epsilon_1 =  3.6 \ 10^{-3}, \ \epsilon_2 =  - 6.6 \ 10^{-3}, \quad \epsilon_3 =  6.7 \ 10^{-3}, \quad \text{for } m_H = 1 \text{ TeV};\\
& \mspace{-5mu} \epsilon_1 =  4.8 \ 10^{-3}, \ \epsilon_2 =  - 7.1 \ 10^{-3}, \quad \epsilon_3 =  6.1 \ 10^{-3}, \quad \text{for } m_H = 300 \text{ GeV};
\end{align}
(these SM contributions are obtained as a linear interpolation from the values listed in \cite{Altarelli:2000ma}).

Fig.~\ref{M1fitb} also reports contours that correspond to several values of the naive unitarity cut-off, \eqref{unitcutoff}. As it can be seen, the model is potentially compatible with EW precision data, even for a relatively small mass scale for the new  heavy vector states. Similarly to the SM case a low Higgs mass is preferred.
Even though a lighter Higgs mass seems to be preferred, the limit it is not as stringent as in the SM: remember that, thanks to the decoupling, in the limit $M_1 \to \infty$ the SM picture is recovered, that is the region on the far right in Fig.~\ref{M1fitb} gives the constraints in the SM case. The main drawback of the model in this limit is that since it has a single extra dimension, which is compact, small and flat, it does not help solving the hierarchy problem: the Higgs mass must still be adjusted through a fine-tuning exactly as in the SM.

\subsection{The model on a slice of AdS$_5$}
\label{RSlimit}

Probably, the most interesting case from the phenomenological point of view is that of an exponentially warped extra-dimension, a slice of $AdS_5$ space. This case corresponds to choosing $b(y) = e^{-2k y}$. The interest of this limit lies both in the possibility of solving the hierarchy problem thanks to an exponential suppression of mass scales on the $y = \pi R$ brane (or IR brane where the Higgs is located) and in the AdS/CFT correspondence \cite{Maldacena:1997re,Witten:1998qj,ArkaniHamed:2000ds,Rattazzi:2000hs}, according to which a model on $AdS_5$ can be viewed as the dual description of a strongly interacting model on four dimensions. In particular, in AdS/CFT fields localized near the IR brane are  interpreted as duals to composite states of the strong sector; in this interpretation the Higgs field is no longer a fundamental field, but only an effective low-energy degree of freedom, just like the KK excitations of the gauge fields.

With this choice we are in a sense come full circle, since we started my theoretical exploration by considering the GD-BESS model, which gives a 4D low-energy effective description of a strongly interacting sector; we generalized that model first to a moose one, then to a 5-dimensional one; finally, thanks to the AdS/CFT correspondence, we can read the generalized 5D model again as an effective description of a strongly interacting theory.

The choice $b(y) = e^{-2k y}$ implies (again by eq. \eqref{doubleint})
\begin{equation}
\label{barMRS}
\frac{1}{\bar{M}^2} = \frac{1}{4k^2} \left(\frac{e^{2k \pi R}(2 k^2 (\pi R)^2 - 2 k \pi R +1) -1}{k \pi R} \right);
\end{equation}
the model has now an extra parameter, the curvature $k$, in addition to the usual $\pi R$, $\bar{g}_5$, $\tilde{g}$, $\tilde{g}'$, $\tilde{v}$ and $\lambda$. Eqs. \eqref{gpar}, \eqref{g1par} and \eqref{vpar} still hold (with the new definition of $\bar{M}$ \eqref{barMRS}); then, after fixing the standard EW input parameters, we are left with three free quantities, $\pi R$ and $\bar{g}_5$ and $k$. Then, if we want this model to be a potential solution to the hierarchy problem, as the RS1 model \cite{Randall:1999ee}, we need to fix the curvature parameter $k$ to be around the Planck scale, $M_P \simeq 10^{19}$ GeV. Then, to have $\bar{M}$ around one TeV, we need $k \pi R \simeq 35$.

Let's look at the spectrum again. In this case, the $C$ and $S$ functions (eq. \eqref{wCS}) are given by:
\begin{equation}
\begin{split}
& S(y,m) = \frac{e^{k y}}{2k} \pi \, m \left(J_1\left(\tfrac{m}{k}\right) Y_1\left(\tfrac{e^{k y} m}{k}\right) - J_1\left(\tfrac{e^{k y} m}{k}\right)Y_1\left(\tfrac{m}{k}\right) \right)\\
& C(y,m) = \frac{e^{k y}}{2 k} \pi \, m \left(J_1\left(\tfrac{e^{k y} m}{k}\right)Y_0\left(\tfrac{m}{k}\right)-J_0\left(\tfrac{m}{k}\right)Y_1\left(\tfrac{e^{k y} m}{k}\right)\right),
\end{split}
\end{equation}
where $J_i$ and $Y_i$ are Bessel function of the first and of the second kind respectively. In this case, not even the condition for the right charged eigenstates can be solved analytically. However, using standard properties of the Bessel functions it is possible to give an estimate for the first eigenvalue,
\begin{equation}
\label{M_1}
M_1 \simeq k \, e^{-k \pi R} \frac{2\sqrt{2}}{\sqrt{4k \pi R - 3}},
\end{equation}
and for the characteristic spacing between two adjacent states, which is approximately constant and equal to $\Delta M = \pi k e^{-k \pi R}$. Notice that
for $k\pi R>>1$ the scale given by $\bar M$ is nothing but $M_1$.

The qualitative analysis made for the flat case generalizes almost verbatim to the AdS case. The main difference is the typical distance between two adjacent eigenstates, which is given by $\Delta M$ rather than simply by $1/R$. In Tables  \ref{spectrum2table} and \ref{spectrum3table}, we show  examples of spectra corresponding to  particular choices of the model parameters. It is interesting to compare this situation to the one of flat case; even though the masses of the first KK level in each sector are roughly the same, the appearance of the second KK level is delayed to a much higher scale.


\begin{center}
\begin{table}
\begin{tabular}{|c|c|c|c|}
\hline
& \textbf{0 mode  (GeV)} & \textbf{$1^{st}$ KK exc.  (GeV)} & \textbf{$2^{nd}$ KK exc.  (GeV)}\\
\hline
\textit{Left charg.} & 80 & 1316 & 16076 \\
\hline
\textit{Right charg.} & - & 1000 & 16054 \\
\hline
\textit{Neutral} & $0,91$ & $1070,1316$ & $16058,16076$ \\
\hline
\end{tabular}
\caption{\small{\textit{Low lying masses  of the spectrum (zero modes and first two KK excitations) for the model in the RS limit, with the following parameter choice: $ k = 6.6 \cdot 10^{18} \emph{ GeV}$, $k\pi R=35$,  $\bar{g}_5 = 1$,
 naive unitarity cut-off equal to $19\cdot 10^{3}$ \emph{GeV}.}}}
\label{spectrum2table}
\end{table}
\end{center}

\begin{center}
\begin{table}
\begin{tabular}{|c|c|c|c|}
\hline
& \textbf{0 mode  (GeV)} & \textbf{$1^{st}$ KK exc.  (GeV)} & \textbf{$2^{nd}$ KK exc.  (GeV)}\\
\hline
\textit{Left charg.} & 80 & 1307 & 8672 \\
\hline
\textit{Right charg.} & - & 1000 & 8632 \\
\hline
\textit{Neutral} & $0,91$ & $1067,1307$ & $8640,8672$ \\
\hline
\end{tabular}
\caption{\small{\textit{Low lying masses  of the spectrum (zero modes and first two KK excitations) for the model in the RS limit, with the following parameter choice: $ k = 4.7 \cdot 10^{18} \emph{ GeV}$, $k\pi R=10$,  $\bar{g}_5 = 1$,
 naive unitarity cut-off equal to $34\cdot 10^{3}$ \emph{GeV}.}}}
\label{spectrum3table}
\end{table}
\end{center}

\begin{figure}
\includegraphics[width=0.45\textwidth]{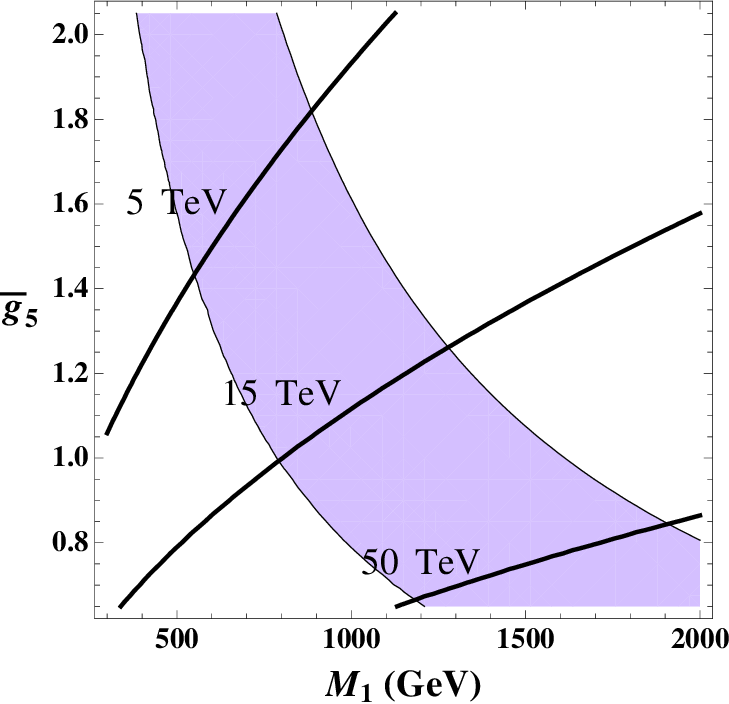}
\includegraphics[width=0.45\textwidth]{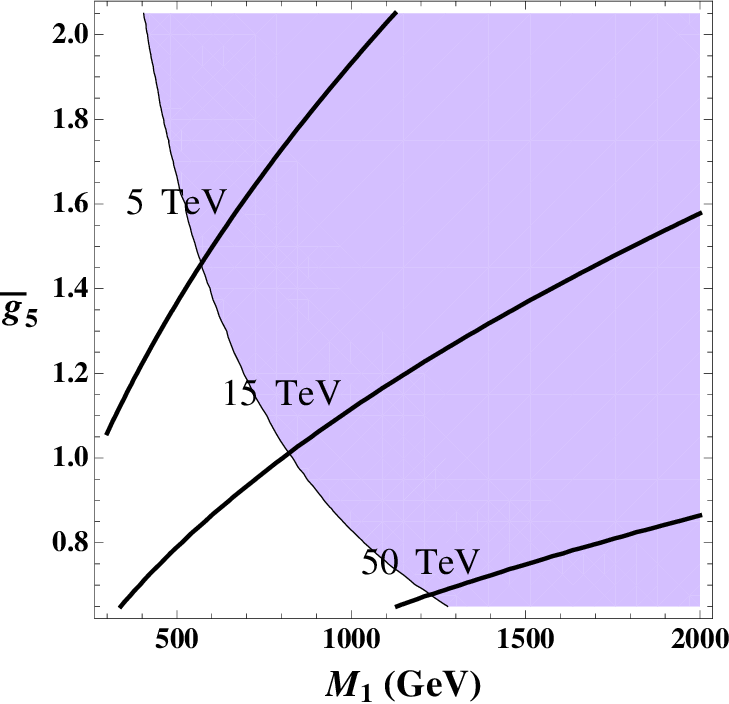}
\caption{\small{\textit{Allowed regions in the $(M_1,\bar{g}_5)$ parameter space for the model in the RS limit ($b(y) = e^{-2k y}$, with $k \pi R$ fixed at $35$), for two values of the Higgs mass: $m_H = 1$ \emph{TeV} (on the left) and $m_H = 300$ \emph{GeV} (on the right), based on electroweak precision constraints. Also shown are the  constraints from naive dimensional analysis (contours correspond to different choices of the  UV cut-off).
 }}}
\label{M1fit}
\end{figure}

Also in this case, we have checked the model against EW precision data using the $\epsilon$ parameters. In Fig.~\ref{M1fit}, we show the allowed region at $95 \%$ C.L. in parameter space $(M_1,\bar{g}_5)$; experimental data and SM radiative correction are the same of the flat case. The regions slightly depend on the choice of $k\pi R$; here we have chosen $k\pi R=35$;  Fig.~\ref{M1fit} also reports contours that correspond to different values of the naive unitarity cut-off, \eqref{unitcutoff} . Notice that in this case, the UV cut-off due to unitarity is generally much lower than it was in the flat case. Nevertheless, the model is again potentially compatible with EW precision data, even when the new heavy vector states have masses around one TeV and an Higgs mass sensibly greater than $100-200$ GeV. The unitarity cut-off scale, which is quite low, calls for an UV extension of the model at an energy scale which is not much higher than the potential reach of the LHC; still the scenario described by the model seems interesting and deserves an accurate study.

The physical content of the 5D D-BESS on an AdS background is very similar to the one of the RS1-like model described in \cite{Carena:2002dz}. In that reference, the authors studied a $SU(2)_L \otimes U(1)_Y$ 5D gauge theory in AdS background, with localized kinetic terms on the IR brane. The main difference between this set-up and the one we have outlined in this work is that we have considered a larger $SU(2)_L \otimes SU(2)_R$ bulk gauge symmetry. Notice, however, that if we add fermions in the simplest way, that is by localizing them on the IR brane (similarly to what was done in refs. \cite{Casalbuoni:2004id,Casalbuoni:2007dk}), then the extra gauge fields (that correspond to what we called the ``right charged sector'') are almost impossible to detect experimentally, since they cannot interact with the fermions (by eq. \eqref{KKBCpiRR} they have no superposition with the IR brane). In fact, as can be seen by the effective Lagrangian calculation of section \ref{lowen5D}, they do not contribute to the $\epsilon$ parameters either. In conclusion, even if the bulk gauge group is different, the phenomenology of the two models is almost identical (the situation changes, however, if fermions are allowed to propagate in the bulk).

This is a very interesting conclusion: working with a completely bottom-up approach, starting from an effective 4D theory - the GD-BESS model - and generalizing, we have arrived at a 5D model that quite closely reproduces a particular version of RS1.

\section{Conclusions}
\label{conclusions}
Among the various TC schemes proposed so far, which have generally
difficulties in satisfying the constraints coming from EW precision
measurements, the Degenerate BESS model \cite{Casalbuoni:1995qt},
provides a low-energy effective scheme which, taking advantage from
a $\left(SU(2) \otimes SU(2)\right)^2$ custodial symmetry,
 leads to a suppressed contribution from the
new physics to the EW  observables. This feature allows new vector
bosons, interpreted as composites of a strongly interacting sector,
at a relatively low energy scale (around a TeV). The interpretation
of the D-BESS  as a four-site "moose" model, makes  the
generalization to $N$ sites natural
\cite{Casalbuoni:2004id,Casalbuoni:2007dk}.

Due to the correspondence at low energy between theories with
replicated 4D gauge symmetries $G$ and theories with a 5D gauge
symmetry $G$ on a lattice, via the deconstruction technique, in this
paper we have considered the continuum limit of the D-BESS model to
a 5D theory. Working in the framework suggested by the AdS/CFT
correspondence, we have considered for the fifth dimension a segment
ending with two branes. The boundary conditions, to be imposed on the
branes, emerge univocally from the deconstruction. The 5D-DBESS
model is described by a $SU(2)_L\otimes SU(2)_R$ gauge theory in the
bulk with boundary kinetic terms, broken both spontaneously and by
boundary conditions.

By choosing the geometry of the fifth dimension as a slice of
AdS$_5$, we find that the 5D-DBESS model can be related to a
realization of the RS1 model \cite{Randall:1999ee} with EW gauge
fields propagating in the bulk and having brane kinetic terms
localized on the infrared brane  \cite{Carena:2002dz}.
The gauge particle content of the theory consists of two charged
sectors, left (including $W$) and right, and one neutral (including
photon and $Z$).

The effective low energy 4D Lagrangian is obtained by means of the
holographic technique. For the purpose to derive the new physics
contribution to the EW observables, it is sufficient to take into
account just the tree-level effects of the heavy resonances that is
to eliminate the bulk fields by their classical equation of motions.
We have derived the expression for the $\epsilon$ parameters and
compared with the experimental values in order to get bounds on the
parameter space of the model. The result is that, also in the
Higgsless case (in which $m_H=1~TeV$ can be considered as a cut-off
of the radiative corrections), some room is still left for low
resonance masses and a weak effective 5D gauge coupling. As a result of the
statistical significance of the 5D-DBESS model grows by lowering
$m_H$. In this respect, and also for pushing forward the unitarity
limit to a scale $\gtrsim 10$ TeV, it is tempting to reintroduce the
Higgs field in our 5D scheme. At least for a particular choice of
the extra-dimensional background, this choice does not force us to
face with the hierarchy problem because, in the ``holographic''
interpretation of AdS$_5$ models
\cite{ArkaniHamed:2000ds,Rattazzi:2000hs}, inspired by the AdS/CFT
correspondence, this Higgs can be seen as a composite state.

The 5D-DBESS on AdS$_5$ then provides a coherent description of the
low energy phenomenology of a new strongly  interacting sector
up to energies significantly beyond the $\sim 2$ TeV limit of the
Higgsless SM, still showing a good compatibility with EW precision
observables.

Two comments are in order:  the  results for the EW precision
parameters are obtained by considering  SM fermions confined to the
infrared brane. However, in this case one generically expects the
emergence of four-fermion operators which induce unacceptable
flavour violations
\cite{Huber:2003tu,Agashe:2004cp,Csaki:2008zd,Casagrande:2008hr,Albrecht:2009xr,Agashe:2009di}.
The well-known cure to this problem is letting fermions propagate in
the bulk. What we  expect is that bulk fermions could  improve  the
compatibility of the model with the present measurements. This will
be the subject of a forthcoming paper.

Another interesting future development is a detailed investigation
on the possibility  of having a heavy Higgs boson, with a mass of
order $300$ GeV or more. Within the SM, global fits indicate that
the Higgs mass cannot be higher than about $160$ GeV; by contrast,
in 5D-DBESS the constraints do not seem so stringent (see figs.
\ref{M1fitb} and \ref{M1fit}) and a heavy Higgs mass could be
allowed.

In conclusion, while the SM Higgs mechanism provides the most
efficient and  economical explanation of the spontaneous electroweak
symmetry breaking, it is still not verified by experiments and it is
not completely satisfactory from a theoretical point of view;
theories in extra dimensions provide a fascinating alternative to
the standard picture, that implies the existence of an interesting
phenomenology that could be observed at LHC.

{\bf Acknowledgments}

The authors would like to thank R. Contino and M. Redi for stimulating discussions and V. Ciulli for clarifying comments on statistical analysis.

\appendix

\newpage

\section{Derivation of the conditions for the KK expansion}
\label{appA}
We will now show how eqs. from \eqref{KKexpL} to \eqref{KKscalarsN} can be derived from the request that the effective 4D Lagrangian is diagonal.
Throughout the following calculation, we will only need to work with the bilinear gauge part of the action \eqref{5Daction}. Expanding the gauge fields as in eq. \eqref{KKexp} \emph{without assuming anything a priori} on the form of the functions $f_j^a$ and $g_j^a$ and the constants $c_j^a$ and carrying out the integration with respect to the extra dimension, we get
\begin{equation}
\label{explag1}
\begin{split}
\lag^{(2)} = -\frac{1}{4} & V_{\mu \nu}^{(j)} V^{(k) \, \mu \nu} A_{j k} - \frac{1}{2} V_\mu^{(j)} V^{(k) \, \mu} B_{j k}\\
-\frac{1}{2} & \de_\mu G^{(j)} \de^\mu G^{(k)} C_{j k} + V_\mu^{(j)} \de^\mu G^{(k)} D_{j k},
\end{split}
\end{equation}
where we defined the matrices:
\begin{equation}
\begin{split}
&  A_{j k} = \frac{1}{g_5^2} \int_0^{\pi R} dy \left(f_{L \, j}^a f_{L \, k}^a + f_{R \, j}^a f_{R \, k}^a \right) + \frac{1}{\tilde{g}^2} f_{L \, j}^a f_{L \, k}^a \big|_{\pi R}\\
& \mspace{50mu} + \frac{1}{\tilde{g}^{'2}} f_{R \, j}^3 f_{R \, k}^3 \big|_{\pi R};\\
&  B_{j k} = \frac{1}{g_5^2} \int_0^{\pi R} dy \, b(y) \left(\de_y f_{L \, j}^a \de_y f_{L \, k}^a + \de_y f_{R \, j}^a \de_y f_{R \, k}^a \right)\\
& \mspace{50mu} + \frac{\tilde{v}^2}{4} b(\pi R) \left(f_{L \, j}^a f_{L \, k}^a + f_{R \, j}^3 f_{R \, k}^3 - 2 f_{L \, j}^3 f_{R \, k}^3 \right) \big|_{\pi R};\\
&  C_{j k} = \frac{1}{g_5^2} \int_0^{\pi R} dy \, b(y) \left(g_{L \, j}^a g_{L \, k}^a + g_{R \, j}^a g_{R \, k}^a \right) + c_j^a c_k^a;\\
&  D_{j k} = \frac{1}{g_5^2} \int_0^{\pi R} dy \, b(y) \de_y \left(f_{L \, j}^a g_{L \, k}^a + f_{R \, j}^a g_{R \, k}^a \right)\\
& \mspace{50mu} - \frac{\tilde{v}}{2} \left. \left(f_{L \, j}^a c_k^a - f_{R \, j}^3 c_k^3 \right) \right|_{\pi R}.
\end{split}
\end{equation}
In the expanded Lagrangian \eqref{explag1}, it is possible to recognize vector and scalar kinetic-like terms, vector mass-like terms and vector / would-be goldstone mixings. However, all those terms are in general not diagonal with respect to the KK number. This is of course a direct consequence of the general nature of the expansion \eqref{KKexp}. However, if the expanded theory is to be consistent, it must be possible to obtain the actual physical degrees of freedom - with explicitly diagonal mass and kinetic terms - by defining appropriate linear combinations of the modes $V_\mu^{(j)}$ and $G^{(j)}$. We then introduce a still general basis change in field space:
\begin{equation}
\label{tilde}
V_\mu^{(j)} = R_{j k} \, \tilde{V}_\mu^{(k)}; \quad G^{(j)} = S_{j k} \, \tilde{G}^{(k)},
\end{equation}
and \emph{require} the Lagrangian \eqref{explag1} to be diagonal in terms of the new degrees of freedom $\tilde{V}_\mu^{(j)}$ and $\tilde{G}^{(j)}$. This means that the matrices $R^T A R$, $R^T B R$, $S^T C S$ and $R^T D S$ (all the fields are real, so we can choose the matrices $R$ and $S$ to be orthogonal) have to be diagonal. Since in general it is not possible to diagonalize four independent matrices using just two rotations, we will need to impose a set of \emph{consistency conditions} on the wave-functions $f_{L, R \, j}^a$ and $g_{L, R \, j}^a$ and the constants $c_j^a$, that will determine the wave functions uniquely.

Let's define:
\begin{equation}
f_{L, R \, j}^a = R_{j k} \, \tilde{f}_{L, R \, k}^a; \quad g_{L, R \, j}^a = S_{j k} \, \tilde{g}_{L, R \, k}^a, \quad c_j^a = S_{j k} \, \tilde{c}_k^a;
\end{equation}
the conditions that we need to impose on the KK modes are then:
\begin{subequations}
\label{masseig}
\begin{align}
\label{masseigA}
\begin{split}
& \frac{1}{g_5^2} \int_0^{\pi R} dy \left(\tilde{f}_{L \, j}^a \tilde{f}_{L \, k}^a + \tilde{f}_{R \, j}^a \tilde{f}_{R \, k}^a \right) +  \frac{1}{\tilde{g}^2} \tilde{f}_{L \, j}^a \tilde{f}_{L \, k}^a \big|_{\pi R}\\
& + \frac{1}{\tilde{g}^{'2}} \tilde{f}_{R \, j}^3 \tilde{f}_{R \, k}^3 \big|_{\pi R} = a_j\delta_{j k};
\end{split}\\
\label{masseigB}
\begin{split}
& \frac{1}{g_5^2} \int_0^{\pi R} dy \, b(y) \left(\de_y \tilde{f}_{L \, j}^a \de_y \tilde{f}_{L \, k}^a + \de_y \tilde{f}_{R \, j}^a \de_y \tilde{f}_{R \, k}^a \right)\\
& + \frac{\tilde{v}^2}{4} b(\pi R) \left(\tilde{f}_{L \, j}^a \tilde{f}_{L \, k}^a + \tilde{f}_{R \, j}^3 \tilde{f}_{R \, k}^3 - 2 \tilde{f}_{L \, j}^3 \tilde{f}_{R \, k}^3 \right)\big|_{\pi R} = b_j \delta_{j k};
\end{split}\\
\label{masseigC}
& \frac{1}{g_5^2} \int_0^{\pi R} dy \, b(y) \left(\tilde{g}_{L \, j}^a \tilde{g}_{L \, k}^a + \tilde{g}_{R \, j}^a \tilde{g}_{R \, k}^a \right) + \tilde{c}_j^a \tilde{c}_k^a = c_j \delta_{j k};\\
\label{masseigD}
\begin{split}
& \frac{1}{g_5^2} \int_0^{\pi R} dy \, b(y) \left(\de_y \tilde{f}_{L \, j}^a \tilde{g}_{L \, k}^a + \de_y \tilde{f}_{R \, j}^a \tilde{g}_{R \, k}^a \right)\\
& \mspace{50mu} - \frac{\tilde{v}}{2} \left. \left(\tilde{f}_{L \, j}^a \tilde{c}_k^a - \tilde{f}_{R \, j}^3 \tilde{c}_k^3 \right) \right|_{\pi R} = d_j \delta_{j k}.
\end{split}
\end{align}
\end{subequations}

We want to reduce the set of eqs. \eqref{masseig} to a more explicit form. As a first thing, consider the integral appearing in the left-hand side of eq. \eqref{masseigB}.  It can be rewritten
\begin{equation}
\label{integralB}
\begin{split}
\mspace{-10mu} \int_0^{\pi R} & b(y) \, \de_y \tilde{f}_{L \, j}^a \de_y \tilde{f}_{L \, k}^a \, dy + \ (L \to R)\\
\mspace{-10mu} = - \int_0^{\pi R} \tilde{f}_{L \, j}^a \de_y (b(y) & \, \de_y \tilde{f}_{L \, k}^a) \, dy + b(\pi R) \, \tilde{f}_{L \, j}^a \de_y \tilde{f}_{L \, k}^a \big|_0^{\pi R} + \ (L \to R).
\end{split}
\end{equation}
If the eigenfunctions $\tilde{f}_{L, R \, j}^a$ satisfy the equation of motion:
\begin{equation}
\label{KKeom1}
\hat{D} f_{L, R \, j}^a = - m_j^2 f_{L, R \, j}^a,
\end{equation}
where we leave the eigenvalue $m_j$ for now unspecified, then the integral in eq. \eqref{integralB} can be further simplified to
\begin{equation}
- m_k^2 \int_0^{\pi R} \tilde{f}_{L \, j}^a \tilde{f}_{L \, k}^a \, dy + b(\pi R) \, \tilde{f}_{L \, j}^a \de_y \tilde{f}_{L \, k}^a \big|_0^{\pi R} + \ (L \to R).
\end{equation}

Now notice from eqs. \eqref{masseig} that the left and right wave-functions only mix through their $3^{rd}$ isospin components. So the conditions \eqref{masseig} receive three separate contributions, one from left wave-functions with isospin $a = 1, 2$, another from $a = 1, 2$ right wave-functions and the last one from mixed left/right $a = 3$ modes. The simplest, most natural choice is to diagonalize the three contributions independently. In this way, we will get three different sets of BCs, that is three decoupled towers of mass eigenstates. While this may not be the most general  solution to eqs. \eqref{masseig}, it is consistent with the symmetry breaking pattern. The general expansion \eqref{KKexp} can then be recast into a more explicit form:
\begin{equation}
\label{KKexp2}
\begin{alignedat}{2}
& W_{L \; \mu}^{1,2} (x,y) = \sum_{n = 0}^\infty f_{L \, n}^{1,2} (y) \, W_{L \, \mu}^{1,2 \, (n)}(x), \quad && W_{L \; 5}^{1,2} (x,y) = \sum_{n = 0}^\infty g_{L \, n}^{1,2} (y) \, G_L^{1,2 \, (n)}(x),\\
& W_{R \; \mu}^{1,2} (x,y) = \sum_{n = 0}^\infty f_{R \, n}^{1,2} (y) \, W_{R \, \mu}^{1,2 \, (n)}(x), && W_{R \; 5}^{1,2} (x,y) = \sum_{n = 0}^\infty g_{R \, n}^{1,2} (y) \, G_R^{1,2 \, (n)}(x),\\
& W_{L \; \mu}^3 (x,y) = \sum_{n = 0}^\infty f_{L \, n}^3 (y) \, N_\mu^{(n)}(x), && W_{L \; 5}^3 (x,y) = \sum_{n = 0}^\infty g_{L \, n}^3 (y) \, G_N^{(n)}(x),\\
& W_{R \; \mu}^3 (x,y) = \sum_{n = 0}^\infty f_{R \, n}^3 (y) \, N_\mu^{(n)}(x), && W_{R \; 5}^3 (x,y) = \sum_{n = 0}^\infty g_{R \, n}^3 (y) \, G_N^{(n)}(x),\\
& \pi^{1,2} (x) = \sum_{n = 0}^\infty c_n^{1,2} \, G^{(n)}(x), && \pi^3 (x) = \sum_{n = 0}^\infty c_n^3 \, G^{(n)}(x).
\end{alignedat}
\end{equation}
As a consequence of this redefinition, the equation of motion \eqref{KKeom1} can also be more explicitly rewritten as three separate equations:
\begin{align}
& \hat{D} f_{L \, n}^{1,2} = - m_{L \, n}^2 f_{L \, n}^{1,2},\\
& \hat{D} f_{R \, n}^{1,2} = - m_{R \, n}^2 f_{R \, n}^{1,2},\\
& \hat{D} f_{L, R \, n}^3 = - m_{N \, n}^2 f_{L, R \, n}^3,
\end{align}
to emphasize the fact that to each sector corresponds a different set of eigenvalues. These three equations reproduce precisely eq. \eqref{KKeomL}, \eqref{KKeomR} and \eqref{KKeomN}.

To go on, assume that the wave-functions obey orthogonality conditions:
\begin{equation}
\label{orthon}
(f_{L \, m}^a,f_{L \, n}^a)_{\tilde{g}} = \delta_{mn}, \quad \frac{1}{g_5^2}(f_{R \, m}^{1,2},f_{R \, n}^{1,2})_{L^2} = \delta_{mn}, \quad (f_{R \, m}^3,f_{R \, n}^3)_{\tilde{g}'} = \delta_{mn},
\end{equation}
where the $(\cdot,\cdot)_{\tilde{g}}$ scalar product was defined in eq. \eqref{dotproL}. With this assumption, the left-hand side of eq. \eqref{masseigA} becomes diagonal, and the equation itself is satisfied by choosing $a_n \equiv 1$. Furthermore, eq. \eqref{masseigB} splits into three independent conditions:
\begin{align}
& \begin{split}
b_n^L \delta_{m n} = & - m_{L \, n}^2 \, \delta_{mn} + \left( \frac{m_{L \, n}^2}{\tilde{g}^2} + \frac{\tilde{v}^2}{4} b(\pi R) \right) \tilde{f}_{L \, m}^{1,2} \tilde{f}_{L \, n}^{1,2} \big|_{\pi R}\\
& + b(\pi R) \left(\tilde{f}_{L \, m}^{1,2} \de_y \tilde{f}_{L \, n}^{1,2}\right)\big|_{\pi R}^0,
\end{split}\\
& \begin{split}
b_n^R \delta_{m n} = & - m_{R \, n}^2 \, \delta_{mn} + \left( \frac{m_{R \, n}^2}{\tilde{g}^2} + \frac{\tilde{v}^2}{4} b(\pi R) \right) \tilde{f}_{R \, m}^{1,2} \tilde{f}_{R \, n}^{1,2} \big|_{\pi R}\\
& + b(\pi R) \left(\tilde{f}_{R \, m}^{1,2} \de_y \tilde{f}_{R \, n}^{1,2}\right)\big|_{\pi R}^0,
\end{split}\\
& \begin{split}
b_n^N \delta_{m n} = & - m_{N \, n}^2 \, \delta_{mn} + \frac{m_{N \, n}^2}{\tilde{g}^2} \left(\tilde{f}_{L \, m}^3 \tilde{f}_{L \, n}^3 + \tilde{f}_{R \, m}^3 \tilde{f}_{R \, n}^3 \right) \big|_{\pi R}\\
& + \frac{\tilde{v}^2}{4} b(\pi R) \left(\tilde{f}_{L \, m}^3 \tilde{f}_{L \, n}^3 + \tilde{f}_{L \, m}^3 \tilde{f}_{L \, n}^3 -2 \tilde{f}_{L \, m}^3 \tilde{f}_{L \, n}^3 \right)\big|_{\pi R}\\
& + b(\pi R) \left(\tilde{f}_{L \, m}^3 \de_y \tilde{f}_{L \, n}^3 + \tilde{f}_{R \, m}^3 \tilde{f}_{R \, n}^3 \right)\big|_{\pi R}^0,
\end{split}
\end{align}
which are identically satisfied as soon as the $f_{L, R n}^a$ obey the BCs \eqref{KKBC0L}, \eqref{KKBCpiRL}, \eqref{KKBC0R}, \eqref{KKBCpiRR}, \eqref{KKBC0N} and \eqref{KKBCpiRN}. Notice that eq. \eqref{KKeomL}, \eqref{KKeomR}, \eqref{KKeomN} together with the above mentioned BCs guarantee the orthogonality of the wave-functions that we assumed in eq. \eqref{orthon}, so we have a self-consistent solution of eqs. \eqref{masseigA} and \eqref{masseigB}. To complete the diagonalization and finally get an expanded bilinear Lagrangian, we just need to solve the last two equations in the set \eqref{masseig}. This can be obtained by imposing the conditions \eqref{KKscalarsL}, \eqref{KKscalarsR} and \eqref{KKscalarsN} respectively on the scalar profiles of the three sectors.


\end{document}